\documentclass[twocolumn]{aa}

\usepackage{graphicx}
\usepackage{txfonts}
\begin{document}
   \title{Multiple stellar populations in Magellanic Clouds clusters.
    \thanks{Based
on observations with the NASA/ESA {\it Hubble Space Telescope},
obtained at the Space Telescope Science Institute, which is operated
by AURA, Inc., under NASA contract NAS 5-26555.}
}
   \subtitle{I. An ordinary feature for intermediate age globulars in
     the LMC?   }

  \author{ A. \,P. \,Milone\inst{1},
        L.\ R.\ Bedin\inst{2},
        G.\ Piotto\inst{1},
        and J.\ Anderson\inst{2}
}
   \offprints{A.\ P.\ Milone}

\institute{
Dipartimento di Astronomia, Universit\`a di
  Padova, Vicolo dell'Osservatorio 3, Padova, I-35122, Italy
\and
Space Telescope Science Institute, 3700 San Martin Drive,
Baltimore, MD 21218, USA
             }

   \date{Received Xxxxx xx, xxxx; accepted Xxxx xx, xxxx}
%__________________________________________________________________
%
\abstract{ 

{\it Context:} The  discovery of multiple main sequences  (MS) in the
massive clusters  NGC~2808 and  Omega Centauri, and  multiple subgiant
branches  in  NGC~1851  and  NGC~6388  has  challenged  the  long-held
paradigm that globular clusters consist of simple stellar populations.
This  evolving   picture  has  been  further   complicated  by  recent
photometric   studies   of    the   Large   Magellanic   Cloud   (LMC)
intermediate-age clusters, where the main sequence turn-off (MSTO) was
found to be bimodal (NGC~1806 and NGC~1846) or broadened (NGC~1783 and
NGC~2173).

{\it Aims:} We have undertaken a study of archival $HST$ images of Large
and  Small Magellanic  Cloud clusters  with the  aim of  measuring the
frequency  of clusters  with evidence  of multiple  or  prolonged star
formation  events and  determining  their main  properties.  We  found
useful images  for 53
 clusters  that cover a  wide range of  ages.  In
this paper,  we analyse the Color-Magnitude Diagrams  (CMD) of sixteen
intermediate-age ($\sim$ 1-3 Gyr) LMC clusters.

{\it Methods:} The data were  reduced by using the method developed by
Anderson  et al.   (2008) and  the photometry  has been  corrected for
differential reddening (where required).  We find that eleven clusters
show an anomalous spread (or  split) in color and magnitude around the
MSTO, even  though the other main  features of the CMD  (MS, red giant
branch, asymptotic giant branch)  are narrow and the horizontal branch
(HB) red-clump  well defined.  By  using the CMD  of the stars  in regions
that surround the cluster, we demonstrate that the observed feature is
unequivocally  associated  to the  clusters.   We use  artificial-star
tests to demonstrate that the spread (or split) is not an artifact due
to photometric errors or binaries.

{\it Results:}  We confirm that  two clusters (NGC~1806  and NGC~1846)
clearly exhibit two distinct MSTOs  and observe, for the first time, a
double  MSTO in  NGC~1751.   In these  three  clusters the  population
corresponding to  the brighter MSTO  includes more than  two-thirds of
cluster  stellar  population.  We  confirm  the  presence of  multiple
stellar  populations in  NGC~1783.  Our  photometry  strongly suggests
that the  MSTO of this cluster  is formed by two  distinct branches. In
seven  clusters (ESO057-SC075,  HODGE7, NGC~1852,  NGC~1917, NGC~1987,
NGC~2108,  and NGC~2154) we  observed an  intrinsic broadening  of the
MSTO that may suggest that these clusters have experienced a prolonged
period of star  formation that span a period between  150 and 250 Myr.
The CMDs of IC~2146, NGC~1644, NGC~1652, NGC~1795 and NGC~1978 show no
evidence of spread or bimodality within our photometric precision.  In
summary 70$\pm$25\% of our sample  are not consistent with the simple,
single stellar population hypotesis.  }
%
%  \keywords{globular clusters:}
   \titlerunning{Multiple stellar populations in Large Magellanic Cloud clusters}
   \authorrunning{Milone et al.}
\maketitle
%________________________________________________________________
%

%%%%%%%%%%%%%%%%%%%%%%%%%%
%
\section{Introduction}
%
%%%%%%%%%%%%%%%%%%%%%%%%%%

Nearly all the clusters that  have been resolved into individual stars
exhibit  a color-magnitude  diagram  (CMD) consistent  with the  stars
belonging to  a single, simple  stellar population.  In  recent years,
however,  thanks  to  the   improving  precision  of  instruments  and
techniques (mainly from space),  the discovery of multiple populations
of stars  in stellar clusters is challenging  this traditional picture
and has led to new views on how clusters form and evolve.

Thus far,  photometry has revealed  multiple stellar populations  in a
few Galactic Globular Clusters  (GGCs) i. e.  Omega Centauri (Anderson
1997, Bedin et  al.\ 2004, Piotto et al.\ 2005,  Sollima et al.\ 2007,
Villanova et  al.\ 2007), NGC~2808  (D'Antona et al.\ 2005,  Piotto et
al.\ 2007), NGC~1851 (Milone et al.\ 2008), and NGC~6388 (Piotto 2008), 
also in some  intermediate-age Large  Magellanic Cloud  (LMC) clusters
(Bertelli et al.\  2003, Mackey \& Broby Nielsen  \ 2007 [M07], Mackey
et al.\ 2008 [M08]) 
and in the Small Magellanic Cloud (SMC) cluster NGC~419 (Glatt et al \ 2008a). 
Each  of the above clusters exhibits a different
pattern of  age spread and/or  chemical enrichment.  It is  clear that
the  star-formation  history  differs  from cluster  to  cluster  (see
Piotto\ 2008 for a review).

%%--------------------------------------------------------------
There is  additional photometric evidence that  other clusters exhibit
some  kind of population  multiplicity.  Marino  et al.\  (2008) found
photometric  evidence of  two distinct  stellar populations  among the
red giant branch  (RGB)  stars  of  M4.  Yong  et  al.\  (2008)  find
abundance  variations in  NGC~6752 that  correlate with  the Str$\rm{\ddot{o}}$mgren
$cy$ index,  and show  eight other clusters  that present a  similar $cy$
spread in their giant branches.

In addition,  many GGCs, including those with  no photometric evidence
for  multiple populations,  exhibit large  star-to-star  variations in
their  chemical abundances  (see Gratton  et al.  2004 for  a review).
Almost all GGCs have  homogeneous Fe-peak-element abundances (with the
only exception being Omega Centauri, hereafter $\omega$~Cen), but they
show  a significant  dispersion in  CNO,  Na, Mg,  Al and  $s$-process
elements.  The  pattern  of   chemical  abundances  in  GGCs  must  be
primordial as  it is observed not  only among RGB  or asimptotic giant
branch (AGB)  stars, but  also among the  sub giant branch  (SGB), and
most  importantly among  unevolved  stars on  the MS
(Gratton et al.\ 2001, 2004).   The presence of a well defined pattern
among Na, O, Mg, and Al suggests that most of the GGCs may have hosted
multiple  episodes of  star formation  separated by  few  hundred Myrs
(D'Antona \& Caloi, 2008).

Unfortunately,  in  an  old  population,  such  a  difference  in  age
corresponds to  a difference of few  hundredths of a  magnitude at the
level  of the  MS turn-off  (TO) and  SGB.  At the
moment it is not possible to firmly establish the presence of a spread
in  a  cluster's  MSTO  with  this amplitude  given  the  presence  of
differential  reddening  and   difficulties  in  doing  systematically
accurate crowded-field photometry at this level of precision.

The  families of young  and intermediate-age  massive clusters that
populate  the Large  and the  Small  Magellanic Cloud  
%A-------
%(SMC)
%--------
 offer  us
precious opportunities to  photometrically search for multiple stellar
populations  because, in  the  CMD of  a  1-3 Gyr  old  cluster, an  age
interval of 200-300  Myr corresponds to a magnitude  difference of a few
tenths of a  magnitude.  Two populations with such  a difference could
easily be distinguished even at  the distance of the Magellanic Clouds
(MC).

Indeed, recent  photometric studies of LMC  clusters have demonstrated
that the presence of multiple stellar populations is not a peculiarity
of  the most  massive Galactic  GGCs.   Bertelli et  al.\ (2003)  have
compared a CMD  of NGC~2173 from FORS/VLT data  with Padova models and
suggested  for this  LMC  cluster a  prolonged star-formation  episode
spanning a  period of  about $300$Myr.  In  a recent paper,  M08 found
that the  main-sequence turn  off (MSTO) for  NGC~1783 reveals  a much
larger spread  in color than  can be explained by  photometric errors.
M07 and  M08 also revealed the presence  of a double MSTO  in the rich
intermediate-age  clusters  NGC~1846  and  NGC~1806 of  the  LMC,  and
suggested that their CMDs unveil  the presence of two populations with
an  age difference  of $\sim  300$~Myr.   

Driven by  these results  on
Galactic and LMC clusters, we  have undertaken an analysis of archival
$HST$ images of  LMC and SMC clusters with the  purpose of measuring the
frequency  of clusters  with evidence  of multiple  or  prolonged star
formation.  The clusters that have been analysed cover a wide range of
ages, from  $\sim 10^{6}$ to $\sim  10^{10}$ yr.  Since  the search of
multiple populations is carried out  through analysis of CMDs,
the study of clusters with different ages requires different analysis.

In this  paper, which  is the first  of a  series, we show  our entire
sample of  ACS/WFC CMDs for  forty-seven LMC and SMC  stellar clusters
and present a detailed  study of sixteen intermediate-age LMC clusters
(between $\sim 1$and $\sim 3$Gyr).   The paper is organized as follow.
Section~\ref{sec:obs}  describes  the  data  and  the  data  reduction
techniques.   In Section~\ref{sec:cmds}  we  present the  CMDs of  the
sixteen intermediate-age clusters.  In Section~4, we confirm the split
of the MSTO of NGC~1806 and NGC~1846, and observe, for the first time,
two distinct MSTOs in NGC~1751. We analyse in details the CMD of these
clusters and measure the fraction  of stars belonging to each of their
MSTO  populations.   We  confirm  the  presence  of  multiple  stellar
populations  in NGC~1783  and suggest  that the  MSTO of  this cluster
should be  formed by two distinct  branches.  In Section 5  we note an
anomalous  spread around  the  MSTO of  seven clusters:  ESO057-SC075,
HODGE~7,  NGC~1852, NGC~1917,  NGC~1987, NGC~2108,  and  NGC~2154.  In
Section~6 we demonstrate  that the observed spread in  the MSTO region
of these  clusters must be real,  as it cannot be  due to differential
reddening,  field contamination, photometric errors or binary stars. In
Section~\ref{sec:iso}  we  determine  the  cluster ages  and  the  age
differences among  different stellar  populations in the  same cluster
through  isochrone   fitting.   Finally,  Section~\ref{sec:conclusion}
includes a short summary of our results.

\section{Observation and data reduction}
\label{sec:obs}
%
%%%%%%%%%%%%%%%%%%%%%%%%%%

In order  to investigate the presence of  multiple stellar populations
in Magellanic  Cloud clusters we  searched the MAST  STScI-archive for
$HST$  images collected  with  the  Wide Field  Channel  (WFC) of  the
Advanced  Camera for  Surveys  (hereafter ACS/WFC).   We found  useful
images for  53 clusters  
from GO~9891  (PI: G. F.  Gilmore ),
  GO~10395 (PI: J. S. Gallagher) and GO~10595 (PI: P. Goudfrooij). The
  result of  this search is presented  in Tab.~1 for  47 clusters from
  GO~9891  and GO~10595, while  the GO~10396  data-set  described in
  Table 1 of  Glatt et al. (2008b) (which includes six SMC clusters) 
  will be used in a future paper focused on SMC.

The photometric  reduction of  the ACS/WFC data  has been  carried out
using the software  presented and described in detail  in Anderson et
al.\ (2008). It consists in  a package that analyses all the exposures
of each cluster  simultaneously in order to generate  a single list of
stars for each  field. Stars are measured independently  in each image
by using the best available PSF models from Anderson \& King (2006).

This routine was  designed to work well in  both crowded and uncrowded
fields and it is able to detect almost every star that can be detected
by eye. It takes advantage of the many independent dithered pointings of
each scene and  the knowledge of the PSF  to avoid including artifacts
in the  list. Calibration of  ACS photometry into the  Vega-mag system
has been performed following recipes in Bedin et al.\ (2005) and using
the zero points given in Sirianni et al.\ (2005).

%\begin{center}
\begin{table*}[ht!]
%\centering
\label{tabelladati}
\scriptsize{
\begin{tabular}{llccrllccr}
\hline\hline  ID & DATE & EXPOSURES & FILT & PROGRAM & ID & DATE & EXPOSURES & FILT & PROGRAM\\
\hline
%-------------------------------------------------------------------------------------
\hline
 ESO~057-SC075 &Nov 05 2006 & 55s$+$2$\times$340s & F435W & 10595   & NGC~1806  & Aug 08 2003 &  300s                  & F555W &  9891  \\
           & Nov 05 2006 &   25s$+$2$\times$340s  & F555W & 10595   &           & Aug 08 2003 &  200s                  & F814W &  9891  \\
           & Nov 05 2006 &   15s$+$2$\times$340s  & F814W & 10595   &           & Sep 29 2005 &   90s$+$2$\times$340s  & F435W & 10595  \\
 ESO~121-SC03  & Oct 07 2003 &   330s             & F555W &  9891   &           & Sep 29 2005 &   40s$+$2$\times$340s  & F555W & 10595  \\
           & Oct 07 2003 &   200s                 & F814W &  9891   &           & Sep 29 2005 &    8s$+$2$\times$340s  & F814W & 10595  \\
 HODGE~7   & Oct 07 2003 &  330s                  & F555W &  9891   &    NGC~1846  & Oct 08 2003 &  300s                  & F555W &  9891  \\
           & Oct 07 2003 &  200s                  & F814W &  9891   &	           & Oct 08 2003 &  200s                  & F814W &  9891  \\
 IC~1660   & Aug 13 2003 &   73s                  & F555W &  9891   &	           & Jan 01 2006 &   90s$+$2$\times$340s  & F435W & 10595  \\
           & Aug 13 2003 &   58s                  & F814W &  9891   &	           & Jan 01 2006 &   40s$+$2$\times$340s  & F555W & 10595  \\
 IC~2146   & Jul 07 2003 &  250s                  & F555W &  9891   &	           & Jan 01 2006 &    8s$+$2$\times$340s  & F814W & 10595  \\
           & Jul 07 2003 &  170s                  & F814W &  9891   &	 NGC~1852  & Oct 07 2003 &  330s                  & F555W &  9891  \\
 KRON~1    & Aug 27 2003 &  480s                  & F555W &  9891   &	           & Oct 07 2003 &  200s                  & F814W &  9891  \\
           & Aug 27 2003 &  290s                  & F814W &  9891   &	 NGC~1854  & Oct 07 2003 &   50s                  & F555W &  9891  \\
 KRON~21   & Aug 22 2003 &  480s                  & F555W &  9891   &	           & Oct 07 2003 &   40s                  & F814W &  9891  \\
           & Aug 22 2003 &  290s                  & F814W &  9891   &	 NGC~1858  & Oct 08 2003 &   20s                  & F555W &  9891  \\
 KRON~34   & Aug 12 2003 &  165s                  & F555W &  9891   &	           & Oct 08 2003 &   20s                  & F814W &  9891  \\
           & Aug 12 2003 &  130s                  & F814W &  9891   &	 NGC~1872  & Sep 21 2003 &  115s                  & F555W &  9891  \\
 LYNDSAY~1 & Jul 11 2003 &  480s                  & F555W &  9891   &	           & Sep 21 2003 &   90s                  & F814W &  9891  \\
           & Jul 11 2003 &  290s                  & F814W &  9891   &	 NGC~1903  & Jui 02 2004 &   50s                  & F555W &  9891  \\
 LYNDSAY~38 & Jun 03 2004 &  480s                 & F555W &  9891   &	           & Jui 02 2004 &   40s                  & F814W &  9891  \\
           & Jun 03 2004 &  290s                  & F814W &  9891   &	 NGC~1917  & Oct 07 2003 &  300s                  & F555W &  9891  \\
 LYNDSAY~91 & Aug 24 2003 &  435s                 & F555W &  9891   &	           & Oct 07 2003 &  200s                  & F814W &  9891  \\
           & Aug 24 2003 &  290s                  & F814W &  9891   &	 NGC~1928  & Aug 23 2003 & 330s                   & F555W &  9891 \\
 LYNDSAY~113 & Jun 03 2004 &  480s                & F555W &  9891   &	           & Aug 23 2003 & 200s                   & F814W &  9891 \\
           & Jun 03 2004 &  290s                  & F814W &  9891   &	 NGC~1939  & Jul 27 2003 & 330s                   & F555W &  9891 \\
 LYNDSAY~114 & Aug 07 2003 &  480s                & F555W &  9891   &	           & Jul 27 2003 & 200s                   & F814W &  9891 \\
           & Aug 07 2003 &  290s                  & F814W &  9891   &	 NGC~1943  & Oct 07 2003 &  50s                   & F555W &  9891 \\
 NGC~265   & Aug 24 2003 &  29s                   & F555W &  9891   &	           & Oct 07 2003 &  40s                   & F814W &  9891 \\
           & Aug 24 2003 &  29s                   & F814W &  9891   &	 NGC~1953  & Oct 07 2003 & 115s                   & F555W &  9891 \\
 NGC~294   & Aug 24 2003 & 165s                   & F555W &  9891   &	           & Oct 07 2003 &  90s                   & F814W &  9891 \\
           & Aug 24 2003 & 130s                   & F814W &  9891   &	 NGC~1978  & Oct 07 2003 & 300s                   & F555W &  9891 \\
 NGC~422   & Oct 07 2003 &  73s                   & F555W &  9891   &	           & Oct 07 2003 & 200s                   & F814W &  9891 \\
           & Oct 07 2003 &  58s                   & F814W &  9891   & 	 NGC~1983  & Oct 07 2003 &  20s                   & F555W &  9891 \\
 NGC~602   & Aug 16 2003 &  29s                   & F555W &  9891   &	           & Oct 07 2003 &  20s                   & F814W &  9891 \\
           & Aug 16 2003 &  29s                   & F814W &  9891   & 	 NGC~1987  & Oct 07 2003 &  250s                  & F555W &  9891  \\
 NGC~1644  & Oct 07 2003 &  250s                  & F555W &  9891   &	           & Oct 07 2003 &  170s                  & F814W &  9891  \\
           & Oct 07 2003 &  170s                  & F814W &  9891   & 	           & Oct 18 2006 &   90s$+$2$\times$340s  & F435W & 10595  \\
 NGC~1652  & Oct 07 2003 &  300s                  & F555W &  9891   & 	           & Oct 18 2006 &   40s$+$2$\times$340s  & F555W & 10595  \\
           & Oct 07 2003 &  200s                  & F814W &  9891   & 	           & Oct 18 2006 &    8s$+$2$\times$340s  & F814W & 10595  \\
 NGC~1751  & Oct 07 2003 &  300s                  & F555W &  9891   & 	 NGC~2002  & Aug 23 2003 & 	 20s &	F555W &	9891 \\
           & Oct 07 2003 &  200s                  & F814W &  9891   &	           & Aug 23 2003 & 	 20s &	F814W &	9891 \\
           & Oct 17-18 2006 & 90s$+$2$\times$340s & F435W & 10595   &	 NGC~2010  & Oct 07 2003 & 	 20s &	F555W &	9891 \\
           & Oct 17-18 2006 & 40s$+$2$\times$340s & F555W & 10595   &	           & Oct 07 2003 & 	 20s &	F814W &	9891 \\
           & Oct 17-18 2006 &  8s$+$2$\times$340s & F814W & 10595   &	 NGC~2056  & Aug 08 2003 & 	170s &	F555W &	9891 \\
 NGC~1755  & Aug 23 2003 &   50s                  & F555W &  9891   &	           & Aug 08 2003 & 	120s &	F814W &	9891 \\
           & Aug 23 2003 &   40s                  & F814W &  9891   &	 NGC~2107  & Oct 07 2003 & 	170s &	F555W &	9891 \\
 NGC~1756  & Aug 12 2003 &  170s                  & F555W &  9891   &	           & Oct 07 2003 & 	120s &	F814W &	9891 \\
           & Aug 12 2003 &  120s                  & F814W &  9891   &	 NGC~2108  & Aug 16 2003 &  250s                  & F555W &  9891  \\
 NGC~1783  & Oct 07 2003 &  250s                  & F555W &  9891   &	           & Aug 16 2003 &  170s                  & F814W &  9891  \\
           & Oct 07 2003 &  170s                  & F814W &  9891   &	           & Aug 22 2006 &   90s$+$2$\times$340s  & F435W & 10595  \\
           & Jan 01 2006 &   90s$+$2$\times$340s  & F435W & 10595   &	           & Aug 22 2006 &   40s$+$2$\times$340s  & F555W & 10595  \\
           & Jan 01 2006 &   40s$+$2$\times$340s  & F555W & 10595   &	           & Aug 22 2006 &    8s$+$2$\times$340s  & F814W & 10595  \\
           & Jan 01 2006 &    8s$+$2$\times$340s  & F814W & 10595   &	 NGC~2154  & Oct 08 2003 &  300s                  & F555W &  9891  \\
 NGC~1795  & Aug 09 2003 &  300s                  & F555W &  9891   &	           & Oct 08 2003 &  200s                  & F814W &  9891  \\
           & Aug 09 2003 &  200s                  & F814W &  9891   &	 RETICULUM & Sep 21 2003 & 	330s &	F555W &	9891 \\
 NGC~1801  & Oct 08 2003 &  115s                  & F555W &  9891   &              & Sep 21 2003 &	200s &	F814W &	9891 \\
           & Oct 08 2003 &   90s                  & F814W &  9891   &  & & & & \\
%
%----------------------------------------------------------------------
\hline
\end{tabular}
}
\caption{Description of the data sets.}
\end{table*}

%%%%%%%%%
%%%%%%%%%
\subsection{Selection of a sample of best measured stars}
\label{sec:2.1}
%%%%%%%%%

Stars can be  poorly measured for several reasons:  crowding by nearby
neighbours, contamination by cosmic-rays (CRs) or image artifacts such
as hot pixels or diffraction spikes.   The goal of the present work is
to  clearly identify multi-populations.  For this  purpose we  need to
select the best-measured stars in the field (i.e., those with the lowest
random and systematic errors), but we also need to have a large enough
statistical sample to be able to identify secondary sequences that may
have many fewer members than the primary ones.

The  software  presented by  Anderson  et  al.\  (2008) provides  very
valuable tools to reach this  goal.  In addition to the stellar fluxes
and positions it  provides a number of parameters that  can be used as
diagnostics   of   the   reliability  of   photometric   measurements.
Specifically these are:
\begin{itemize}
\item the  rms  of  the  magnitudes measured  in  different  exposures
  ($rms_{m_{\rm   F435W}}$,  $rms_{m_{\rm  F555W}}$,  and  $rms_{m_{\rm
  F814W}}$, available only for clusters with more than one exposure in the
same filter, cfr. Tab.~1);
\item the  rms of   the  positions measured  in
  different exposures transformed in a common distortion-free
  reference frames ($rms_{\rm X}$, and $rms_{\rm Y}$);
\item the residuals to the PSF fit for each star ($q_{m_{\rm F435W}}$,
  $q_{m_{\rm F555W}}$, and $q_{m_{\rm F814W}}$; this is what Anderson et al.\ 2008,
  define as {\em quality fit});
\item the ratio between the estimated flux of the star in a 0.5 arcsec
  aperture,  and  the flux  from  neighboring  stars  within the  same
  aperture ($o_{m_{\rm  F435W}}$, $o_{m_{\rm F555W}}$,  and $o_{m_{\rm
  F814W}}$, again see Anderson et al.\ 2008 for details).
\end{itemize}
We used these parameters to select a sub-sample of stars with the best
photometry.

\begin{figure}[ht!]
   \centering
   \includegraphics[width=8.5 cm]{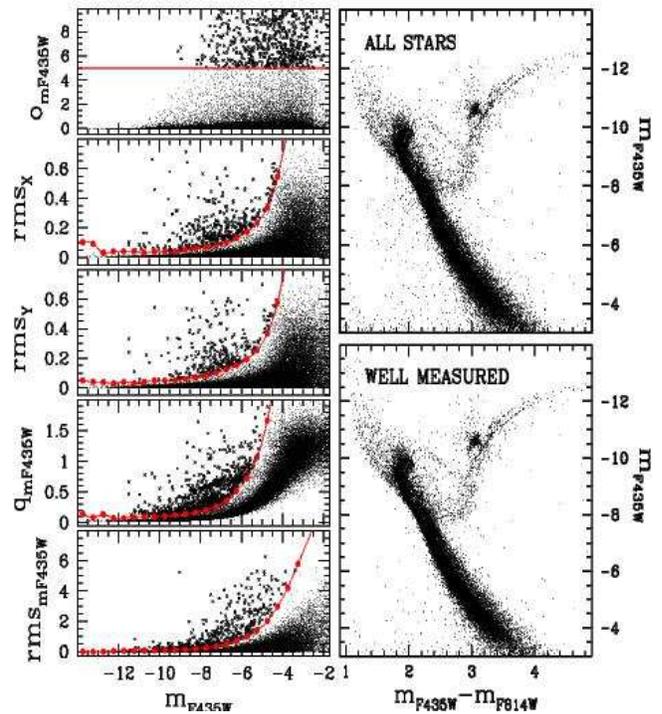}
      \caption{ Diagnostic  parameters used  to select the  stars with
        the  best photometry  are  plotted as  a  function of  $m_{\rm
          F435W}$ for  NGC~1806. Red lines  separate the well-measured
        stars (thin points) from those that are more likely to have a
        poorer photometry (thick points).   On the right we compare the
        CMD of all the measured stars (top) and of stars that pass our
        criteria of selection (bottom).  
      }
         \label{selezioni}
   \end{figure}

In the five left panels  of Fig.~\ref{selezioni}, we show the criteria
that  we  have used  to  select  the sample  of  stars  with the  best
photometry in the F435W band  for NGC~1806.  The photometric system at
this   stage  is   kept   in  instrumental   magnitudes,  $-2.5   {\rm
log}_{10}({\rm flux})$, where the flux is expressed in photo-electrons
recorded in the reference exposure.

We note  a clear trend  of the quality-fit  and $rms$ parameters  as a
function of the magnitude  due to the decreasing signal-to-noise ratio
(S/N).  In  order to  select well-measured stars  at different  S/N we
adopted the following procedure.

We began  by dividing all the  stars into magnitude bins.  The size of
each bin varied from one cluster to another depending on the number of
stars; for  each of them, we  computed the median $rms$  ($q$) and the
$68.27^{\rm  th}$  percentile (hereafter  $\sigma$).   The median  was
derived recursively: after each  computation, all stars exceeding four
times $\sigma$  were provisionally  rejected until the  next iteration
and the median  was recomputed. This procedure was  repeated until two
subsequent measures of  the median differ by less than  the 1\% of the
value.

Finally, we arrived at the  red line of Fig.~\ref{selezioni} by adding
to the  median of each bin $N$  times $\sigma$.  This gave  us the red
circles, which we then fitted with  a spline.  All stars below the red
line in  each plot have been  flagged as well-measured.  The factor $N$
ranges  from  5 to  6,  and  has been  chosen  in  order  to draw  the
boundaries that follow the bulk  of the distribution of each parameter
value.

The neighbour-contamination parameters do  not show a clear trend with
magnitude, so  we simply flagged  as well-measured all the  stars with
$o_{m_{\rm F435W,~F555W,~F814W}}<5$.  

Obviously, the $rms$ of the
magnitudes was  not available  in the case  of clusters with  only one
image  per  band  (HODGE~7,  IC~2146,  NGC~1644,  NGC~1652,  NGC~1795,
NGC~1852,  NGC~1917,  NGC~1978, and  NGC~2154).  In  such cases  those
selections were not applied.

In the right panels of Fig.~\ref{selezioni} we show the CMD of all the
measured  stars of  NGC~1806  (top) and  of  stars that  pass all  the
selection criteria (bottom).
%__________________________________________________________________
%__________________________________________________________________
%
\subsection{Artificial-star tests}
\label{sec:artstars}
%%%%%%%%%

The artificial-star (AS) experiments used  in this paper have been run
following the procedures described in Anderson et al.\ (2008).

First of all,  for each cluster, we produced an  input list with about
$10^5$ stars located on the entire ACS field of view.  It includes the
coordinates of the stars ($X_{\rm in}$, $Y_{\rm in}$) in the reference
frame,  and the magnitudes  in the  F435W (or  F555W) and  F814W bands
($mv_{\rm in}$,  $mi_{\rm in}$ ).   We generated the  artificial stars
with  a  flat   luminosity  function  in  the  F814W   band  and  with
instrumental  magnitudes from  $-$5  to $-$14.   We placed  artificial
stars fainter than the MSTO on the MS ridge line that is determined as
described  in  section \ref{simulation1}  and  added artificial  stars
brighter than the MSTO along the isochrone that best fits the observed
CMD (see section \ref{sec:iso}).

The   program   described   in   Anderson  et   al.\   (2008)   allows
artificial-star  tests to  be performed  for one  star at  a  time and
entirely  in  software guaranting  that  artificial  stars will  never
interfere  with  each  other,  no  matter how  many  tests  are  done.
Therefore it makes very simple  to do artificial star tests because it
is  not necessary  to add  an array  of non-interfering  stars  on the
images, reducing all the images each time.  For each star in the input
list, the  routine adds the star  to each exposure  at the appropriate
place with the  appropriate flux, and measures the  images in the same
manner  as real  stars, producing  the  same output  parameters as  in
section \ref{sec:2.1}.   If the input and the  output positions differ
by  less than  0.5  pixel and  the  fluxes differ  by  less than  0.75
magnitudes, than the artificial star is considered as found.

Artificial stars played a crucial  role in this analysis; they allowed
us to  determine the completeness level  of our sample  and to measure
the fraction  of chance-superposition ``binaries''.   The completeness
depends on the  crowding conditions as well as  on stellar luminosity.
Our procedures account for both of these.
Our  goal  in  the  AS  tests  is  to  probe  how  incompleteness  and
photometric errors vary with crowding or brightness.  So, we generated
a list  of artificial stars  that had colors  that placed them  on the
average cluster sequence and  positions drawn from the overall cluster
radial  distribution.  

\begin{figure*}[ht!]
   \centering \includegraphics[width=\textwidth]{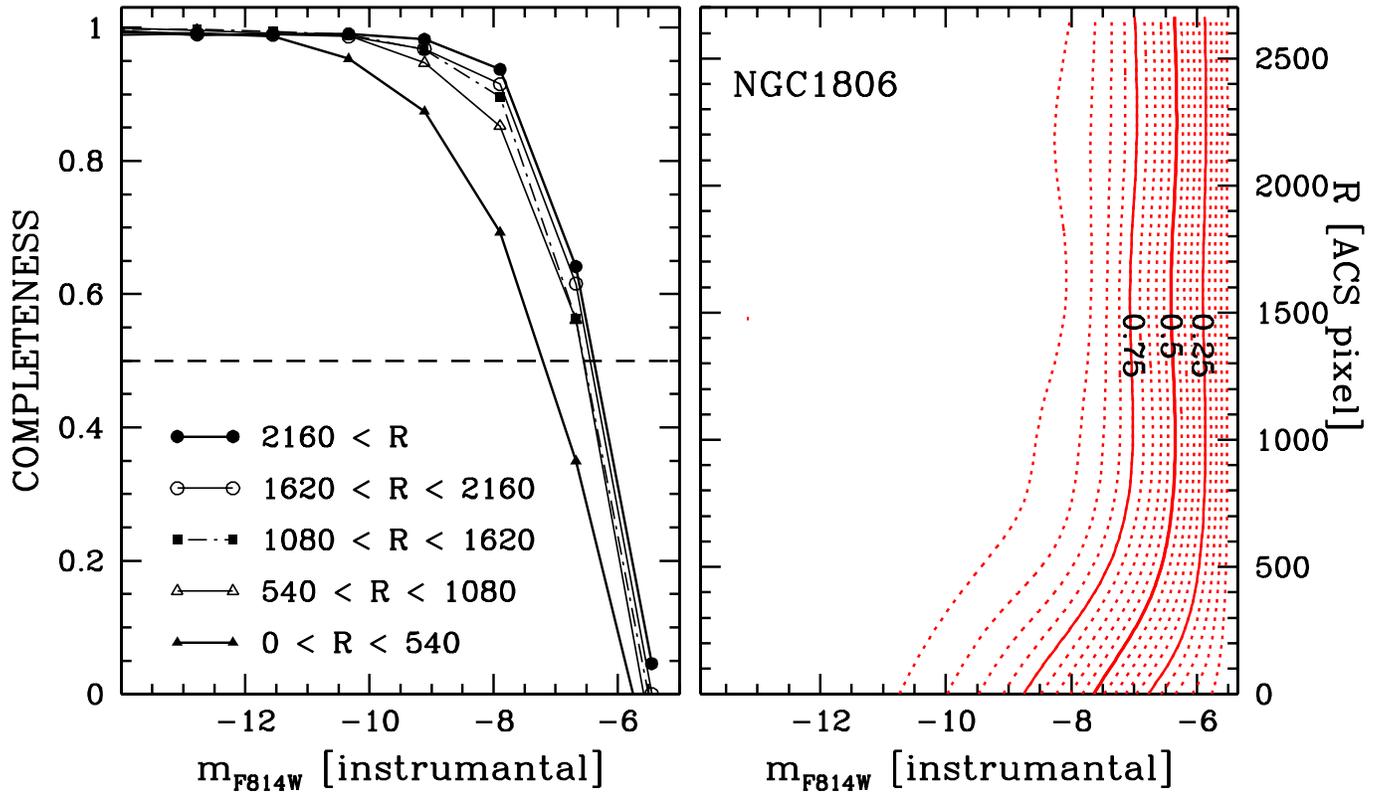}
      \caption{  Left:  Completeness  as  a function  of  the  $m_{\rm
          F814W}$  magnitude  in  five  annuli.   Right:  Completeness
          contours  in  the  radial  distance versus  $m_{\rm  F814W}$
          magnitude plane.  }
         \label{compl}
   \end{figure*}

We divided the ACS/WFC field for each cluster into 5 concentric annuli
centered on  the cluster center,  and within each annulus  we examined
the AS results in eight magnitude  bins, from $-14$ to $-5$.  For each
of  these  5$\times$8 grid  points,  we  then  determined the  average
completeness by taking the ratio  of the recovered to input artificial
stars  within that bin.   This grid  then allowed  us to  estimate the
completeness for any  star at any position within  the cluster.  As an
example, the left panel  of Fig.~\ref{compl} shows the completeness as
a function  of the instrumental  $m_{\rm F814W}$ magnitude in  each of
the five annuli we used to  divide the field of NGC~1806.  Finally, we
interpolated  the  grid  points  and derived  the  completeness  value
associated with  each star. The right panel  of Fig.~\ref{compl} shows
the  completeness  contours  in  the  radius  versus  $m_{\rm  F814W}$
magnitude plane. Continuous lines correspond to completeness levels of
0.25, 0.50 and 0.75. Dotted lines indicate differences of completeness
of 0.05.

%__________________________________________________________________
%__________________________________________________________________

\section{The cluster CMDs}
\label{sec:cmds}
\begin{figure*}[ht!]
   \centering
   \includegraphics[width=\textwidth]{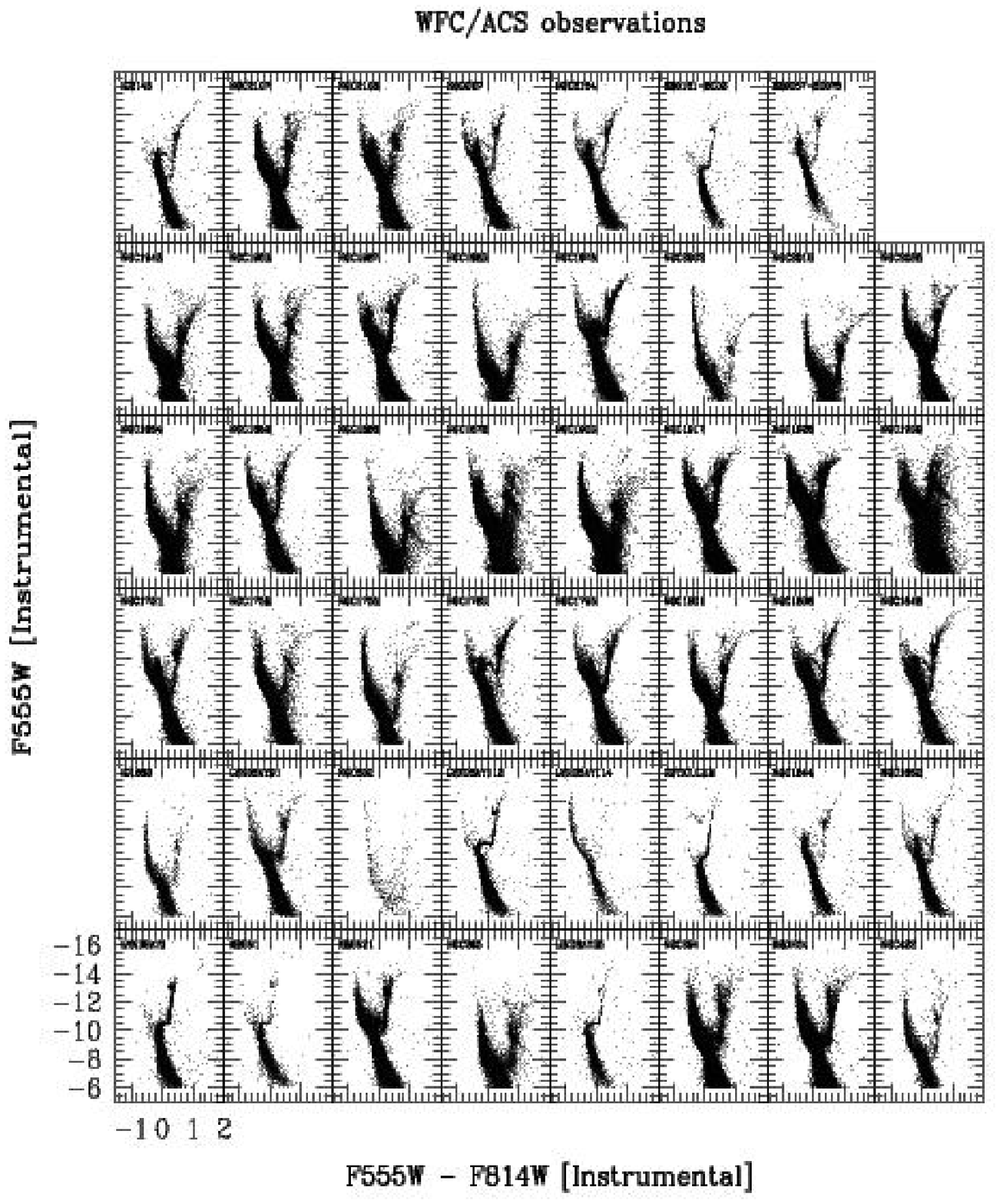}
      \caption{ Instrumental $m_{\rm F555W}$ vs. $m_{\rm F555W}-m_{\rm
          F814W}$ for 47 MCs clusters.  }
         \label{fig:clusters}
   \end{figure*}

%
%%%%%%%%%%%%%%%%%%%%%%%%%%
In  Fig.~\ref{fig:clusters} we show  the instrumental  $m_{\rm F555W}$
versus  $m_{\rm F555W}-m_{\rm  F814W}$  CMDs of  all  the 47  clusters
from GO~9891 and GO~10595
presented in Section~\ref{sec:obs}.  It is clear from this figure that
our sample  contains clusters  at  many different  stages of
evolution.   This  paper  will  deal  with the  CMDs  of  the  sixteen
intermediate-age clusters (from $\sim$ 1 to $\sim$ 3 Gyr). The other
clusters  will  be  covered   in  forthcoming  papers.   All  selected
intermediate age clusters are LMC members.

%__________________________________________________________________
%__________________________________________________________________
Most of the CMDs  plotted in Fig.~\ref{fig:clusters} show considerable
evidence  of contamination,  mainly due  to MC  field stars.   For the
purposes of this paper, it is crucial to determine as well as possible
the  stellar distribution in  the field-star  CMD, since  without such
attention,  field sequences  could  be erroneously  attributed to  the
cluster and interpreted as an additional population.

Since most of the clusters of  our sample cover a small portion of the
ACS/WFC  field  of  view,  we   can  easily  isolate  a  CMD  that  is
representative  of the  field  population surrounding  the cluster  by
selecting the portion of the  ACS/WFC image that are most distant from
the  cluster center,  so  that the  contamination  of cluster  members
should be negligible and, in many cases, almost absent.

In order to  minimize the fraction of field stars  in the cluster CMD,
we selected two  regions with the same area.   The first region, which
below we will call the 'cluster field', is centered on the cluster and
includes the region with the  highest density of cluster members (with
the only exception  of NGC~1978, where we chose  an annulus around the
cluster center  in order  to exclude the  crowded cluster  core).  The
second  region is far  enough from  the cluster  center that  very few
cluster stars would be expected to be present there. We will call this
field 'reference field' and its  stars should be representative of the
typical population in front of  and behind the cluster.  Since cluster
stars in the  vicinity of the MSTO are the main  target of the present
study, we defined  the area of these regions such  that the density of
stars   within  a  magnitude   from  the   TO,  namely   with  $m_{\rm
  F814W}<(m_{\rm F814W}^{\rm TO}+1)$, in the cluster field would be at
least a fixed  number ($N$) of times that of  the reference field.  In
most  cases we could  set $N=5$,  which would  mean that  even without
field correction, we have at  most a 20\% contamination.  An exception
is NGC~1917 which is projected  upon a densely populated region of the
LMC.  In  this case  we used $N=3$.   For the most  populated clusters
(IC~2146, NGC~1806, NGC~1846, and NGC~1978) we adopted $N=10$.

\begin{figure*}[ht!]
   \centering
   \includegraphics[width=\textwidth]{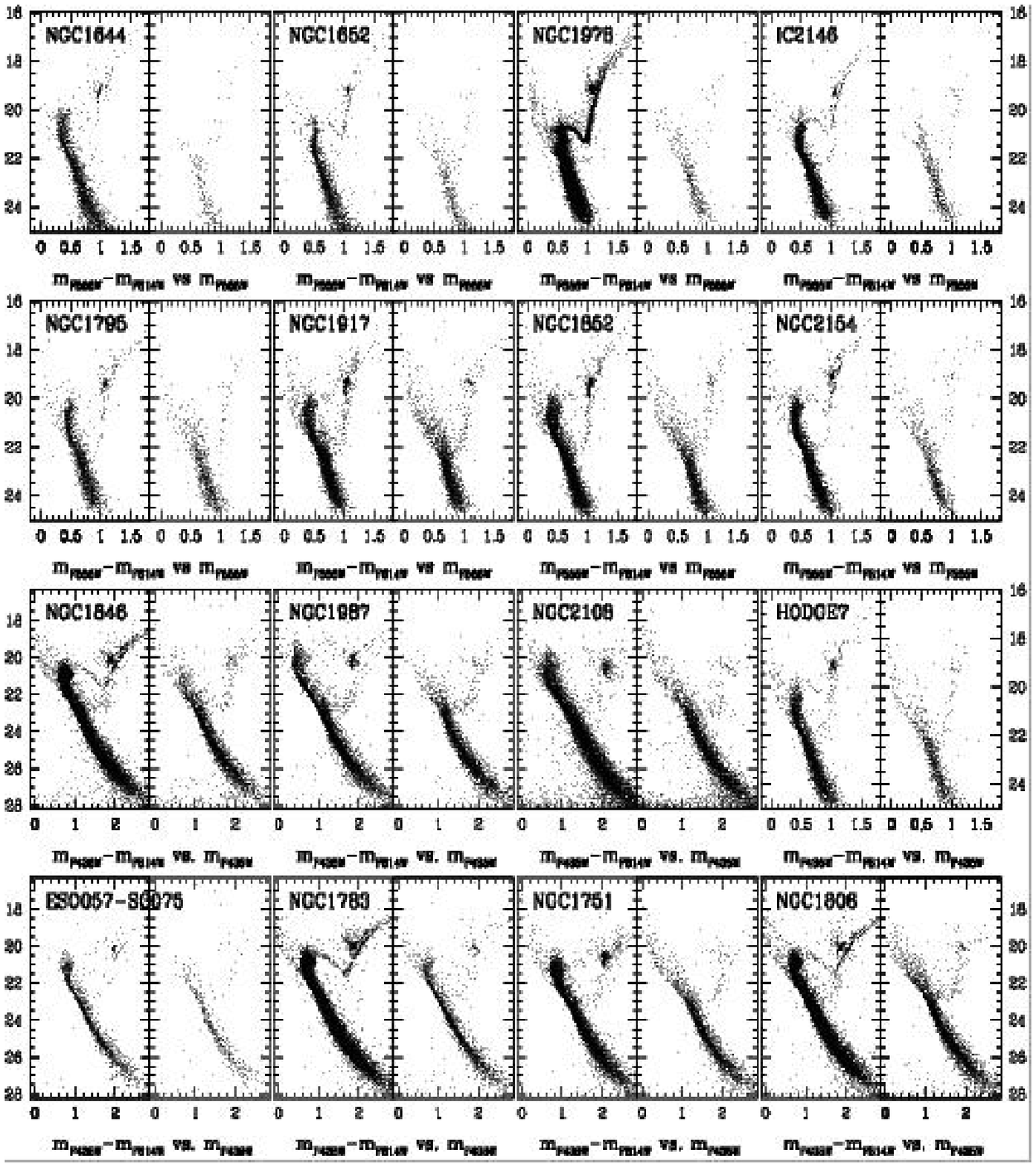}
      \caption{Comparison of the CMD of cluster and
      reference fields for the 16 selected intermediate-age clusters.
       }
         \label{cmds}
   \end{figure*}

In Fig.~\ref{cmds}, we  compare the CMDs of the  cluster and reference
fields  for  the sixteen  intermediate  age  clusters  that have  been
studied in this paper.
%__________________________________________________________________
%__________________________________________________________________
All  the reference-field  CMDs selected  with this  criterion  share a
broadened, young  main sequence and  a 5-6 Gyr old  stellar population
that departs from  the MS at around $m_{\rm  F555W}=22.5$ and populate
the evolved  portions of the  CMD.  In Sec.~\ref{sec:fieldsubtraction}
we will use  the CMDs of the reference field  to decontaminate the CMD
of the cluster.

\subsection{Differential reddening}
It is well known that differential reddening causes a shift of all the
CMD  features parallel  to the  reddening line  and tends  to randomly
broaden them.  A quick look at  the CMDs presented in this paper shows
that most of  the RGBs and AGBs are narrow and  well defined, and that
the HB red-clump  is compact. Moreover, we have  divided the cluster field
into  many subregions  (the exact  number varies  from one  cluster to
another, depending  on the number of  stars) and compared  the CMDs of
stars located in each of them. In most cases, we found no evidence for
an offset  among the CMDs  and therefore, any variations  of reddening
should be negligible.

\begin{figure}[ht!]
   \centering
   \includegraphics[width=9.2cm]{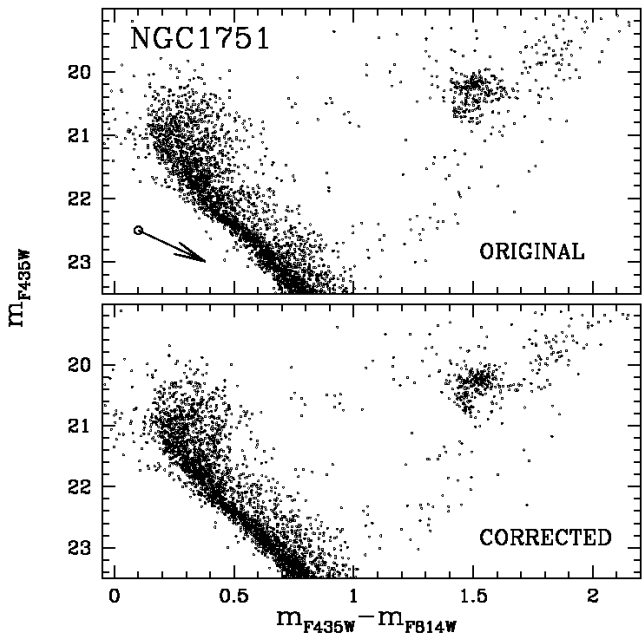}
      \caption{Comparison  between  the CMD  of  NGC~1751 before  (top
      panel) and after (bottom  panel) the correction for differential
      reddening. The arrow indicates the reddening direction.}
         \label{red1751}
   \end{figure}

\begin{figure}[ht!]
   \centering
   \includegraphics[width=9.2cm]{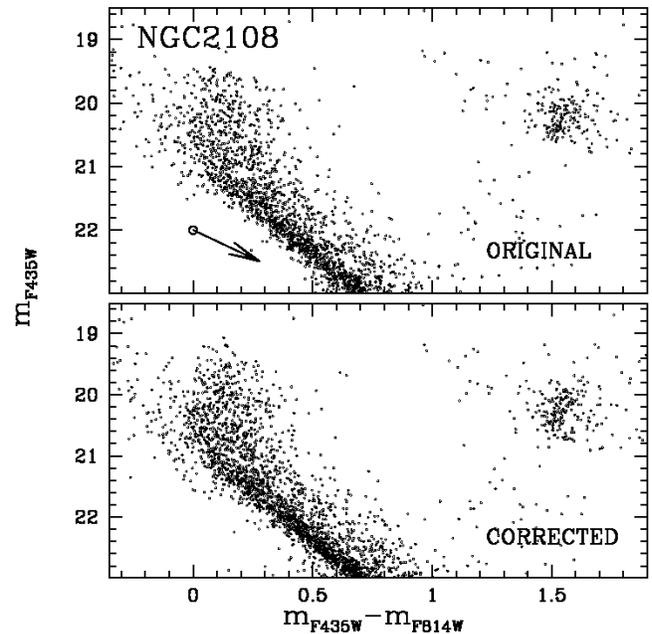}
      \caption{As in Fig.~\ref{red1751} for NGC~2108.}
         \label{red2108}
   \end{figure}

Two exceptions  to this reddening-free rule are  NGC~1751 and NGC~2108.
We show their CMD in Fig.~\ref{red1751} and \ref{red2108}. The confused
shape  of the  HB red-clumps  and of  other primary  features  suggests the
presence of differential reddening.  Since differential reddening acts
along the  direction of the reddening  arrow shown at  the lower left,
sequences  (such as  the SGB)  that are  aligned perpendicular  to the
reddening line are most affected.
%__________________________________________________________________
%
%__________________________________________________________________
%%__________________________________________________________________
%%
%%__________________________________________________________________

For this reason, a search for  a possible split or a spread around the
MSTO  requires  an accurate  correction  of  the  effects produced  by
differential reddening  on the observed  CMD. In order to  correct for
differential  reddening,  we  have  used the  procedure  described  in
Sarajedini  et al.\  (2007).   Briefly: we  define  the fiducial  main
sequence for the cluster and tabulate,  at a grid of points across the
field, how the  observed stars in the vicinity of  each grid point may
systematically lie  to the red or  the blue of  the fiducial sequence;
this systematic  color offset is indicative of  the local differential
reddening.

In the lower panel of Fig.~\ref{red1751}, we show the corrected CMD of
NGC~1751.   It should  be noted  how,  after the  correction has  been
applied, all the main features of the CMD become
narrower and clearly  defined, confirming that most of  the effects of
differential reddening have been  removed.  The improvement of the CMD
is particularly evident for the stars of the HB red-clump; the tightness
along the reddening line of this clump means that
the spread (or split) we see in the MSTO cannot be due to differential
reddening.   Figure~\ref{red2108}  illustrates   the  effects  of  the
reddening correction  in NGC 2108.   The total amount  of differential
reddening  within the cluster  region is  $\Delta$ $\textit {E(F435W-F814W})$=0.10
for NGC~1751 and $\Delta$ $\textit{E(F435W-F814W})$=0.08 for NGC~2108.

The two clusters that suffer of a sizable differential reddening are
not at any particular angular distance from the center of the LMC,
or with respect to the Milky Way, compared to the other forteen clusters.
They might just fall in some gas-dust complex structure poorly known,
and/or with a limited spatial extension. 
%%%%%%%%%%%%%%%%%%%%%%%%%%
%
\section{Clusters with a double MSTO: NGC~1806, NGC~1846 and NGC~1751}
\label{sec:2MSTO}
\begin{figure}[ht!]
   \centering
   \includegraphics[width=8.75 cm]{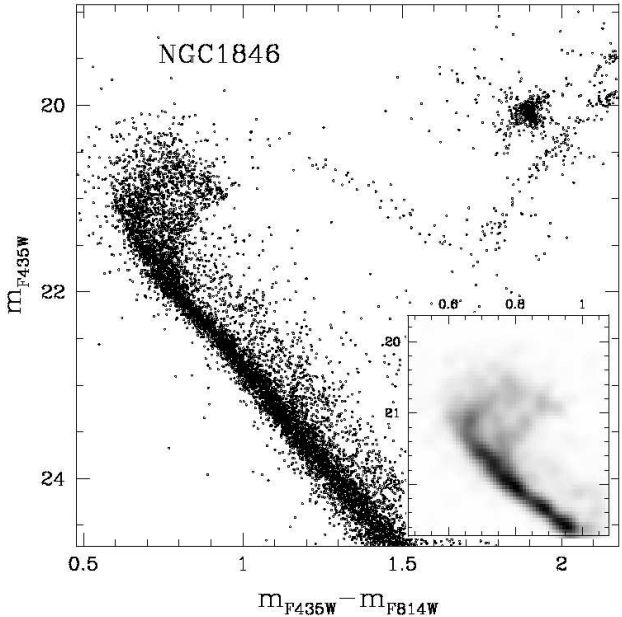}
      \caption{  CMD of NGC~1846.  All detected  sources in  the inner
       field that successfully passed  all the selection criteria have
       been plotted. In the inset we show the Hess diagram for the CMD
       region around the MSTO.  }
         \label{NGC1846}
   \end{figure}

\begin{figure}[ht!]
   \centering
   \includegraphics[width=8.75 cm]{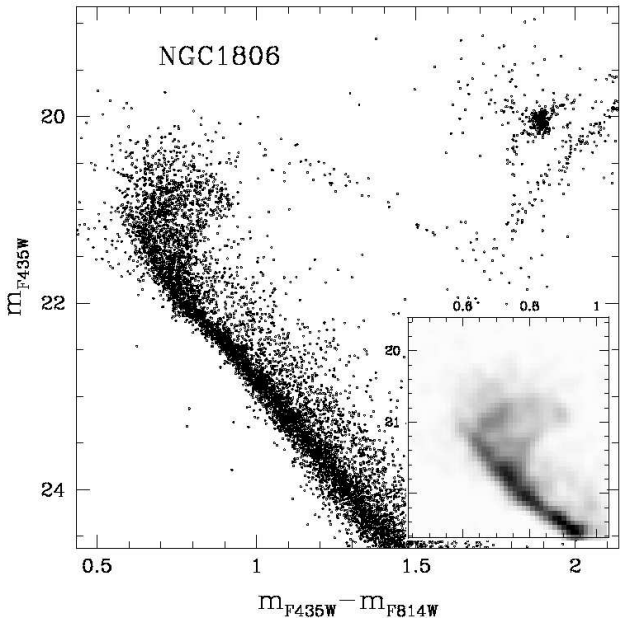}
      \caption{ As in Fig.~\ref{NGC1846} for NGC~1806.
       }
         \label{NGC1806}
   \end{figure}

\begin{figure}[ht!]
   \centering
   \includegraphics[width=8.75cm]{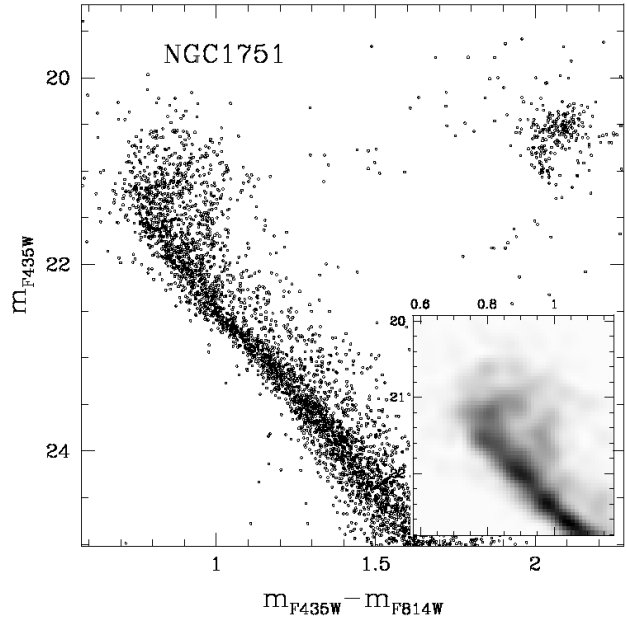}
      \caption{As in Fig.~\ref{NGC1846} for NGC~1751.}
         \label{ngc1751}
   \end{figure}

At  least three  clusters  of  our sample  clearly  show two  distinct
MSTOs:\ NGC~1846, NGC~1806, and NGC~1751.  The split is evident in the
CMDs  of the  cluster  fields that  are  shown in  Fig.~\ref{NGC1846},
\ref{NGC1806} and \ref{ngc1751}, and is exalted by the Hess diagram in
the inset.  The presence of two, distinct TOs in NGC~1806 and NGC~1846
was discovered  by M07 and  M08, who also  found no difference  in the
spatial  distribution between the  stars in  the brighter  and fainter
MSTOs (hereafter bMSTO and fMSTO).
%%__________________________________________________________________
%__________________________________________________________________
%__________________________________________________________________
%__________________________________________________________________
%__________________________________________________________________
%
%__________________________________________________________________
In this work,  we take advantage from the  high photometric quality of
our CMDs  to measure the fraction  of stars belonging  to each turnoff
population. To do this, it is  necessary to select and compare the two
groups  of  bMSTO  and  fMSTO.

\begin{figure*}[ht!]
   \centering
   \includegraphics[width=\textwidth]{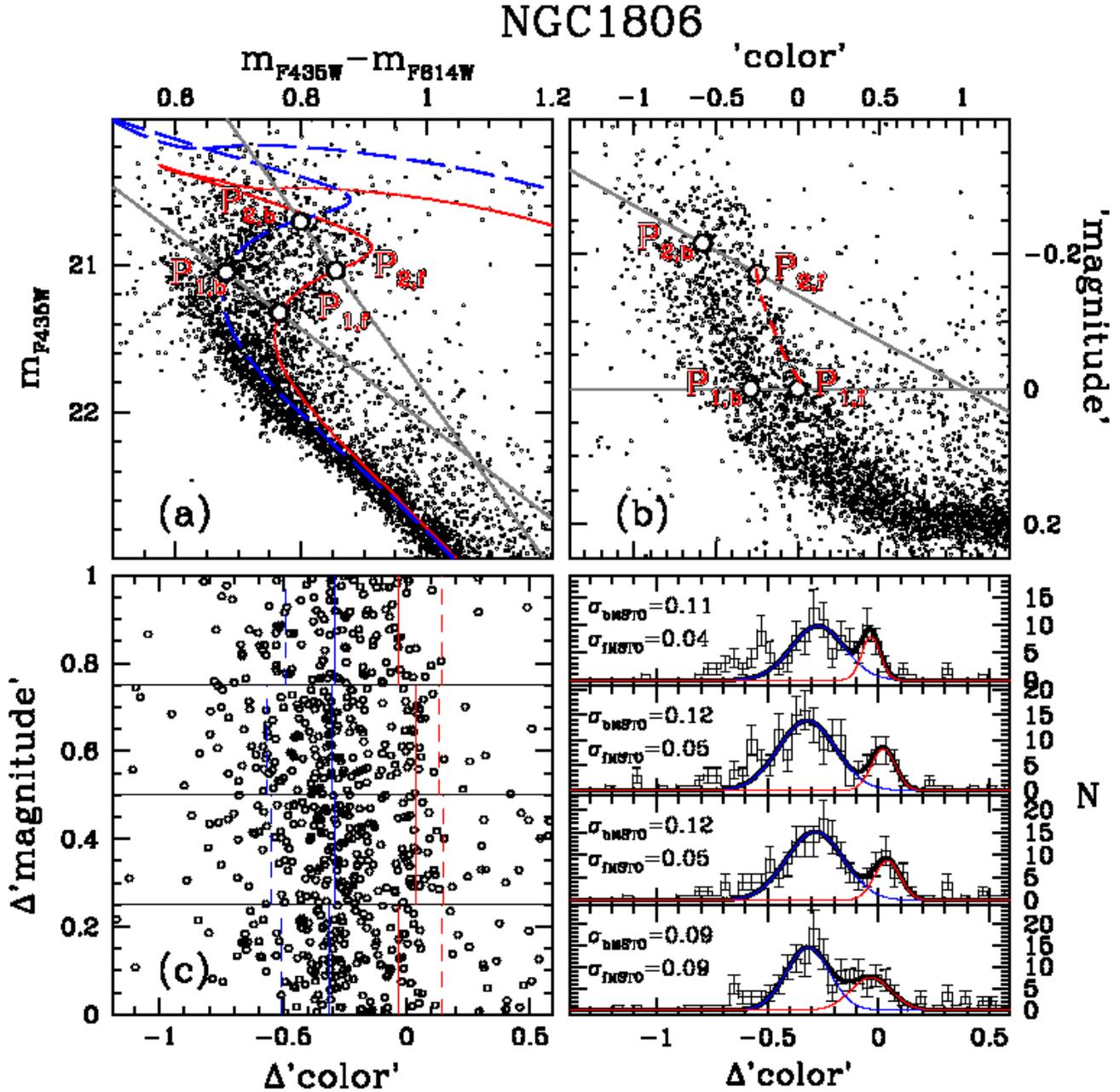}
      \caption{This  figure  illustrates   the  procedure  adopted  to
        measure the fraction of stars belonging to the bMSTO and fMSTO
        in  NGC~1806.   Panel  (a)  shows  a  zoom  of  the  CMD  from
        Fig.~\ref{NGC1806}  with  the isochrones  that  best fits  the
        fMSTO and  the bMSTO overimposed.  The grey  lines delimit the
        portion of the CMD where the split is more evident. Only stars
        from this region are used to measure the population ratio.  In
        Panel (b) we shifted and  rotated the reference frame of Panel
        (a). Red dashed line is  the fiducial of the region around the
        fMSTO.  In Panel  (c) we plotted stars between  the grey lines
        but after the subtraction of  the 'color' of the region around
        the  fMSTO fiducial  from the  'color'  of each  star and  the
        division of the 'magnitude' of each star by the 'magnitude' of
        the upper  grey line.  The  four right bottom panels  show the
        $\Delta$'color'    distribution     for    stars    in    four
        $\Delta$'magnitude'   bins.  The   solid  lines   represent  a
        bigaussian  fit. For  each bin,  the dispersions  of  the best
        fitting gaussians are indicated.  }

         \label{popRATIO1806}
   \end{figure*}

This procedure is illustrated in Fig.~\ref{popRATIO1806} for NGC~1806.
We  used the  CMD  of the  cluster  field, with  foreground/background
contamination  removed  as described  in  Section  6.1.   We began  by
finding the  isochrones that  went through the  two sequences  and the
giant   branch   (as   it    will   be   described   in   details   in
section~\ref{sec:iso}).   These are the  red and  the blue  lines.  We
then  defined by  hand two  pairs of  reference points:  $P_{1,f}$ and
$P_{2,f}$ on  the fMSTO isochrone  and $P_{1,b}$ and $P_{2,b}$  on the
bMSTO isochrone.  The two  pairs of points  have been chosen  with the
criterion of delimiting the region of  the CMD where the split is more
evident and have been used to  draw the grey lines in panel (a).  Only
stars contained in  the region between these lines  have been used for
the following analysis.

In panel  (b), we  have shifted and  rotated the reference  frame such
that the  new origin  corresponds to $P_{1,f}$  and the  abscissa goes
from  $P_{1,b}$ to $P_{1,f}$.   For simplicity,  in the  following, we
will refer  to the  abscissa and ordinate  of this reference  frame as
'color'  and 'magnitude'.   The red  dashed  line is  a fiducial  line
through the region  close to the fMSTO.  We drew it  by marking on the
region close to  the fMSTO four points, equally  spaced in 'magnitude'
and drawing, a line through them by means of a spline fit.

In panel  (c) the 'color'  of this line  has been subtracted  from the
color of each star, and the  'magnitude' of each star has been divided
by the  'magnitude' of a stars with  the same 'color' that  lie on the
line that goes from $P_{1,f}$ to $P_{1,b}$.

For the analysis  that follows, we have divided  the $\Delta$'magnitude' range
into $N$ bins.  We used $N=4$  for NGC~1806 and NGC~1846 and $N=2$ for
the less populated NGC~1751.  In  each of them, we have determined the
fraction of stars  belonging to each MSTO as follows.   Our aim was to
model  the $\Delta$'color'  distribution  by fitting  the  sum of  two
partially overlapping  gaussians, but we need to  reduce the influence
of outliers (such as stars  with poor photometry, residual field stars
and binaries).  To this end, we did a preliminary fit of the gaussians
using all available stars, then we rejected all the stars distant more
than two $\sigma_b$ from the bMSTO and less than 2 $\sigma_f$ from the
fMSTO and repeated the fit (where the $\sigma$'s are those of the best
fitting  gaussians in  each $\Delta$'magnitude'  bin fitted  to the  fMSTO and
bMSTO respectively).

The  continuous vertical lines  show the  centers of  the best-fitting
gaussians in  each $\Delta$'magnitude' interval.  The  red dashed line
is located  two $\sigma_{f}$ on the  "blue" side of the  fMSTO and the
blue line runs two $\sigma_{b}$ on the "red" side of the bMSTO,

%__________________________________________________________________
%

It  is important  to notice  here that,  to each  point ($P_{1,f(b)}$,
$P_{2,f(b)}$) that we have  arbitrarely selected on the two isochrones
with the only purpose of  disentangle the two SGBs, corresponds a mass
($\mathcal{M}_{P1f(b)}$,  $\mathcal{M}_{P2f(b)}$).   In order  to  obtain  a more  correct
measure  of  the fraction  of  stars in  each  of  the two  population
($f_{bMSTO}$, $f_{fMSTO}$) it  must be noted that we  are dealing with
two different mass  intervals ($\mathcal{M}_{P2f}-\mathcal{M}_{P1f} \neq \mathcal{M}_{P2b}-\mathcal{M}_{P1b}$)
and we have to compensate for two facts: first, more massive stars are
more rare, and second, more massive stars evolve faster.

We calculated the fraction of stars in each branch as:
\begin{center}
$f_{bMSTO}  =  \frac { \frac {A_{b}} {N_{b}/N_{f}}}  {A_{f} +  \frac
    {A_{b}} {N_{b}/N_{f}}} $ \\
$f_{fMSTO}  =  \frac {A_{f}} {A_{f} +  \frac {A_{b}} {N_{b}/N_{f}}} $
\end{center}
where $A_{b}$ and $A_{f}$ are the  area of the gaussians that best fit
the bMSTO and  the fMSTO, and $N_{f(b)}=\int_{P_{1,f(b)}}^{P_{2,F(b)}}
\phi(\mathcal{M})d\mathcal{M}$, being adopting for $\phi(\mathcal{M})$
the Salpeter (1955) IMF.

We find that 74$\pm$4 \% of  stars of NGC~1806 belong to the bMSTO and
26$\pm$4\% to the fMSTO.  In the  case of NGC~1846 we have 75$\pm$3 \%
of stars in the bMSTO and 25$\pm$3 \% in the fMSTO.  Finally, 69$\pm$4
\% of  the NGC~1751 stars belong to  the bMSTO and the  31$\pm$4 \% to
the  fMSTO.  Interestingly  enough,  a  similar population  ratio
between the  bright and  the faint  SGB ratios has  been found  in the
Galactic globular clusters NGC 1851 (Milone et al.\ 2008) and NGC 6388
(Piotto et al, in preparation).  We note  here that D'Antona  and Coloi  (2008) predict
that  more than  50\% of  the cluster  stars must  be coming  from the
second  (younger) population  in  their intermediate  mass AGB  ejecta
pollution scenario  proposed to  explain multiple populations  in star
clusters.

\subsection{The (double ?) MSTO of NGC~1783.}
\label{sec:1783}
%__________________________________________________________________
   \begin{figure}[ht!]
   \centering
   \includegraphics[width=8.75 cm]{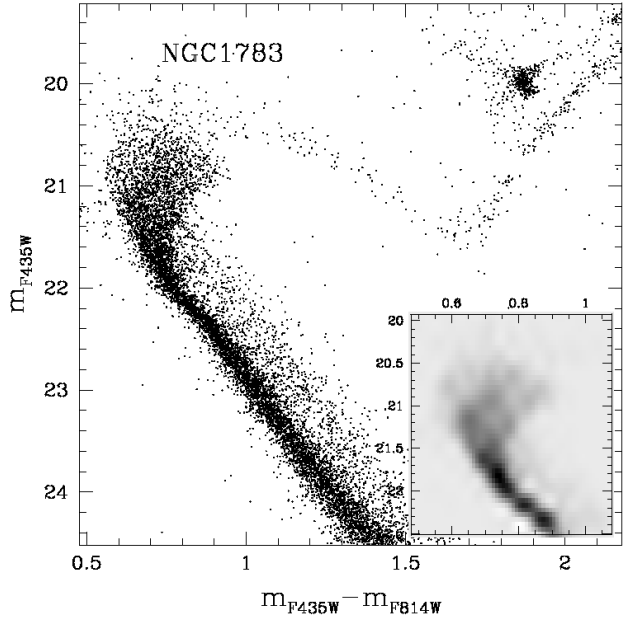}
      \caption{As in Fig.~\ref{NGC1846} for NGC~1783.
       }
         \label{NGC1783}
   \end{figure}
%__________________________________________________________________
A  spread in  color around  the MSTO  of NGC~1783  was first  noted by
Mucciarelli  et  al.\  (2007).   Unfortunately their  $m_{\rm  F555W}$
versus $m_{\rm F555W}-m_{\rm F814W}$ CMD  had  low photometric  accuracy,
because  it was obtained  from the  GO~9891 images  alone.  Therefore,
they were unable to distinguish  between the intrinsic spread in color
and  the  broadening   expected  by  photometric  uncertainties.   M08
obtained a CMD for this cluster  from data with higher S/N (using both
GO~10595 and  GO~9891 images) and  demonstrated that NGC~1783  shows a
much larger spread in color than what would be expected by photometric
errors alone.

Our  CMD  of  NGC~1783  is  shown in  Fig.~\ref{NGC1783}  and  clearly
confirms the anomalous  spread around the MSTO. In  addition, the Hess
diagram in  the inset  reveals a splitted  MSTO and  strongly suggests
that the  apparent spread could be  attributed to the  presence of two
distinct branches which are closely  spaced and poorly resolved by the
observations.

\section{Possible evidence of prolonged star formation}
\label{sec:spMSTO}
%__________________________________________________________________
%
   \begin{figure} [ht!]
   \centering
   \includegraphics[width=8.75cm]{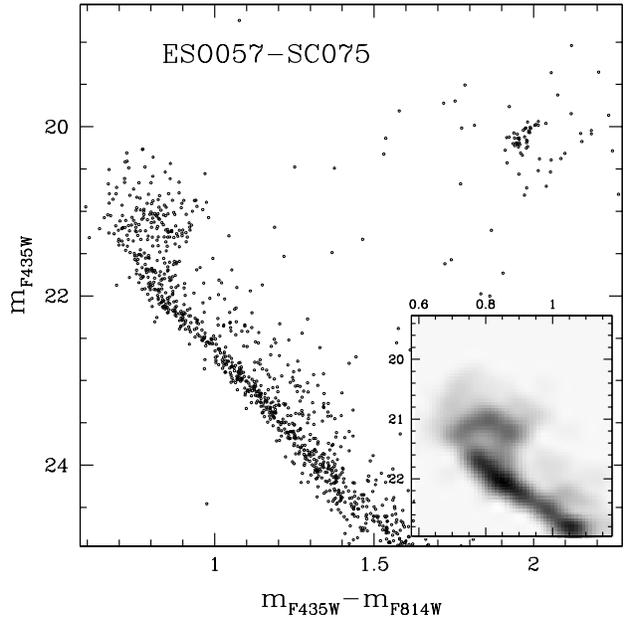}
      \caption{As in Fig.~\ref{NGC1846} for ESO057-SC075.}
         \label{eso}
   \end{figure}
%__________________________________________________________________
%__________________________________________________________________
%
   \begin{figure}[ht!]
   \centering
   \includegraphics[width=8.75cm]{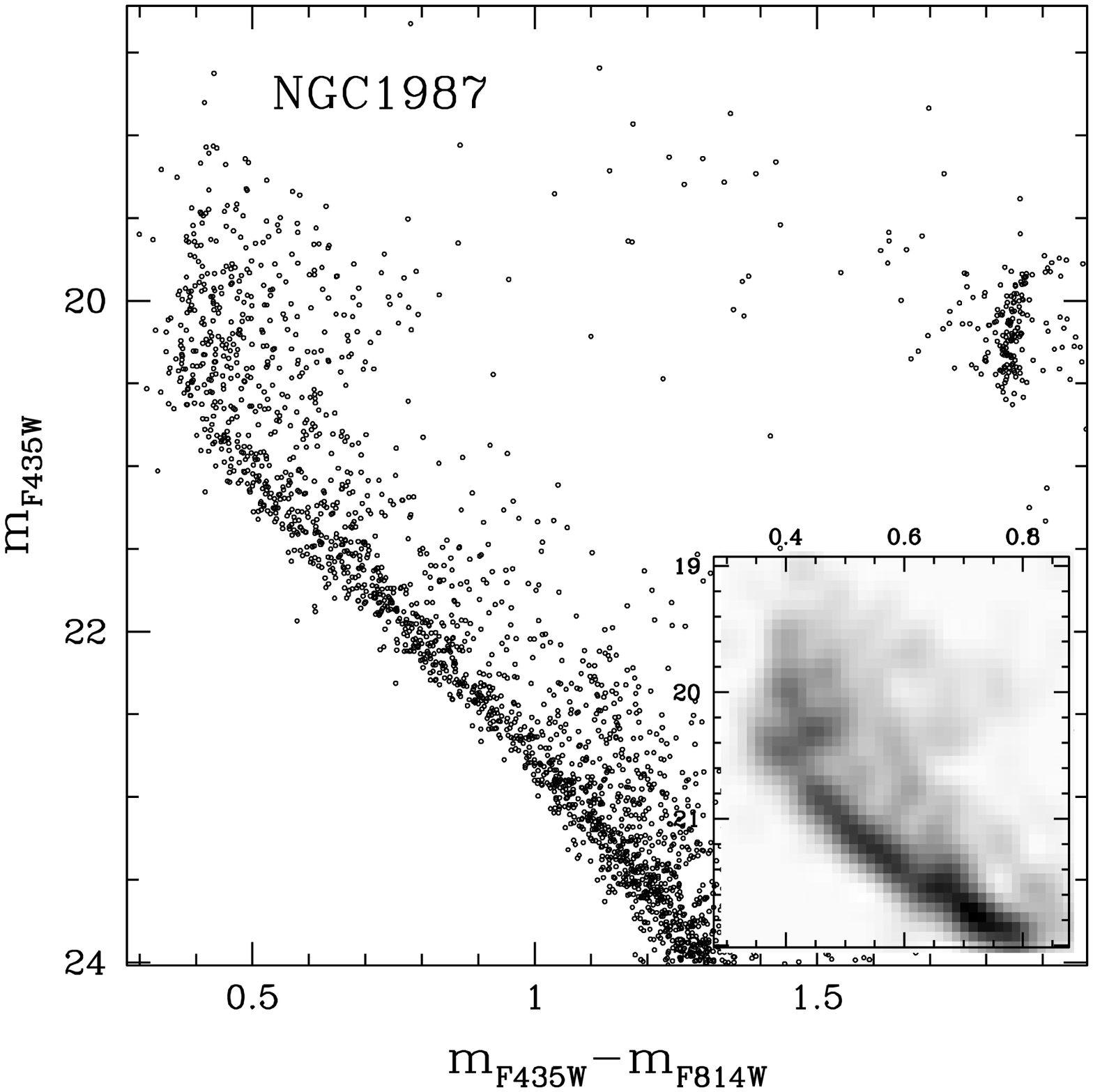}
      \caption{As in Fig.~\ref{NGC1846} for NGC~1987.}
         \label{ngc1987}
   \end{figure}
%__________________________________________________________________
%__________________________________________________________________
%
   \begin{figure}[ht!]
   \centering
   \includegraphics[width=8.75cm]{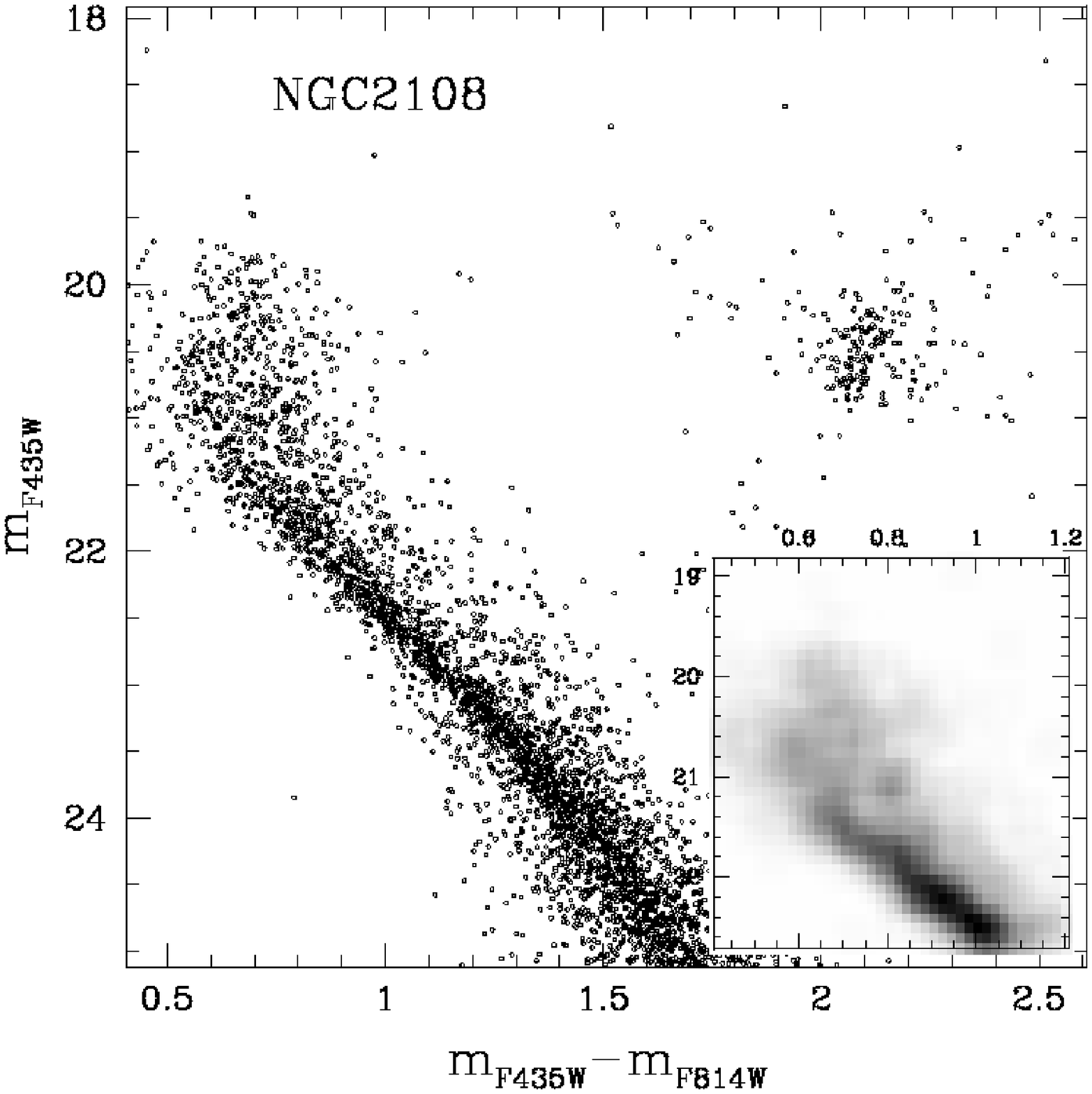}
      \caption{As in Fig.~\ref{NGC1846} for NGC~2108.}
         \label{ngc2108}
   \end{figure}
%__________________________________________________________________
%__________________________________________________________________
%
   \begin{figure}[ht!]
   \centering
   \includegraphics[width=8.75cm]{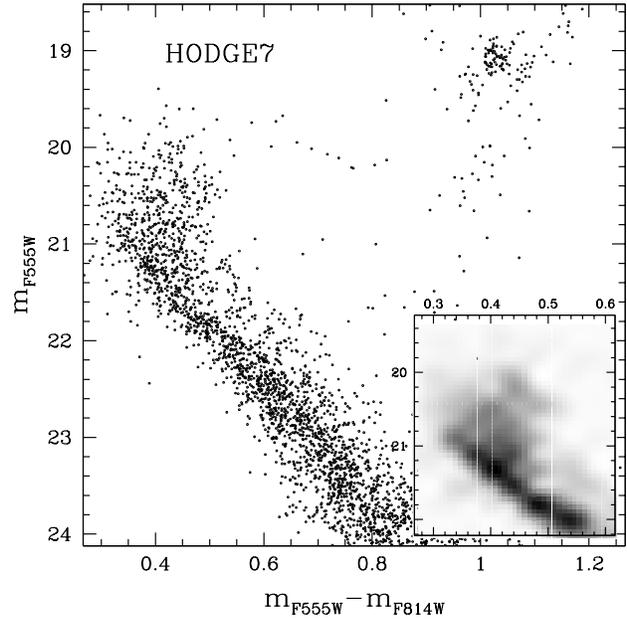}
      \caption{As in Fig.~\ref{NGC1846} for HODGE~7.}
         \label{hodge7}
   \end{figure}
%__________________________________________________________________
%__________________________________________________________________
%
   \begin{figure}[ht!]
   \centering
   \includegraphics[width=8.75cm]{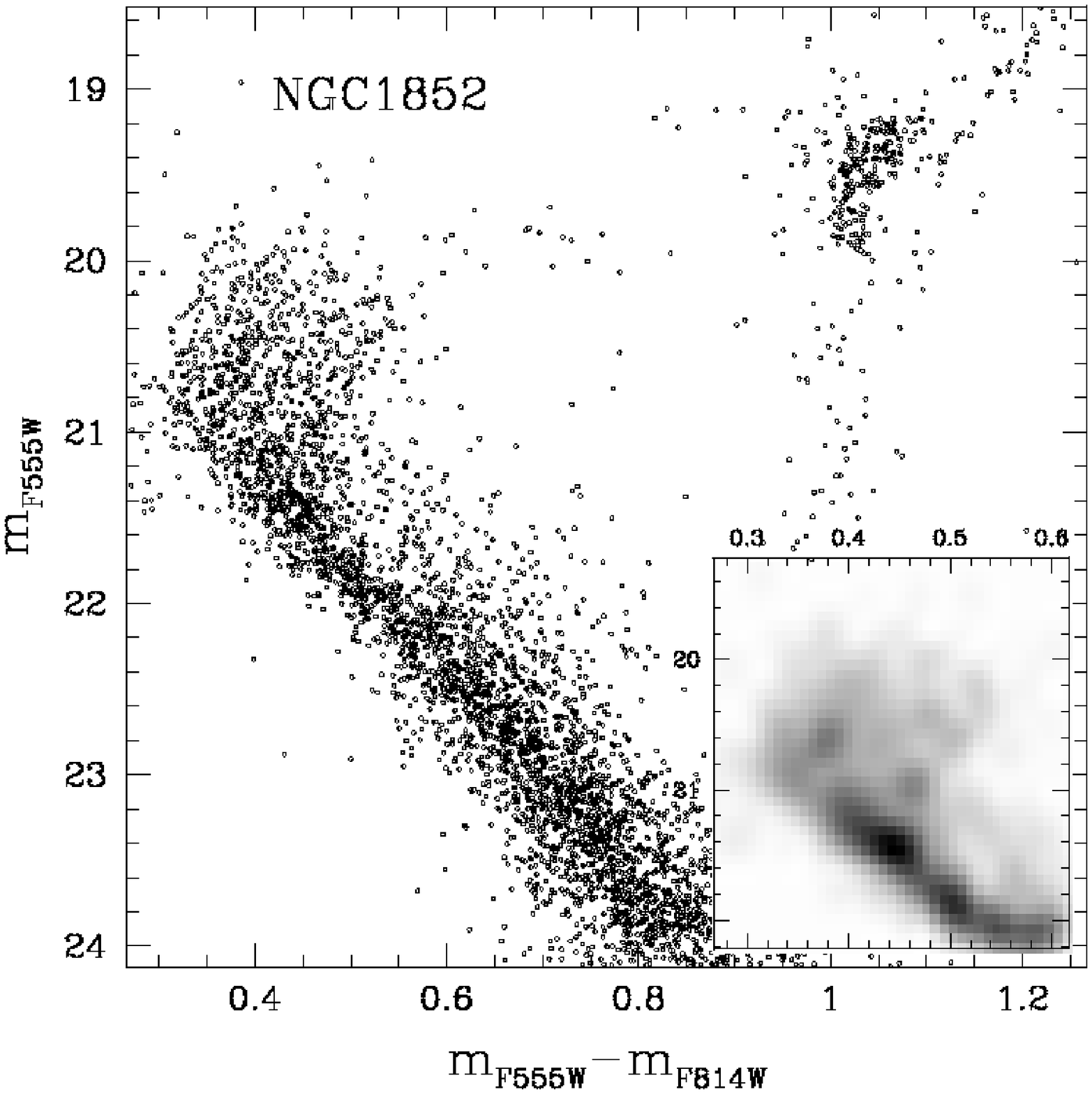}
      \caption{As in Fig.~\ref{NGC1846} for NGC~1852.}
         \label{ngc1852}
   \end{figure}
%__________________________________________________________________
%__________________________________________________________________
%
   \begin{figure}[ht!]
   \centering
   \includegraphics[width=8.75cm]{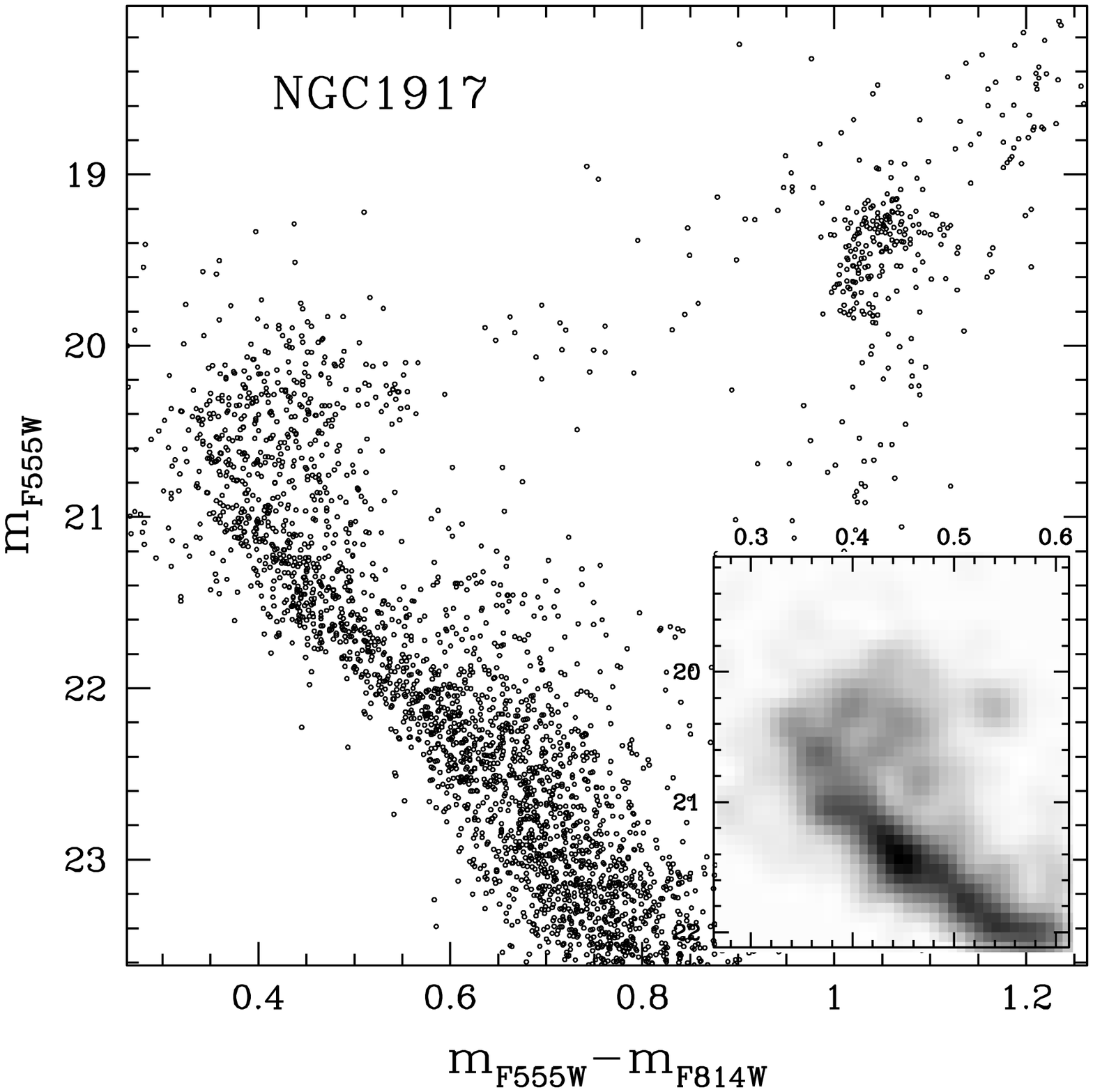}
      \caption{As in Fig.~\ref{NGC1846} for NGC~1917.}
         \label{ngc1917}
   \end{figure}
%__________________________________________________________________
%__________________________________________________________________
%
   \begin{figure}[ht!]
   \centering
   \includegraphics[width=8.75cm]{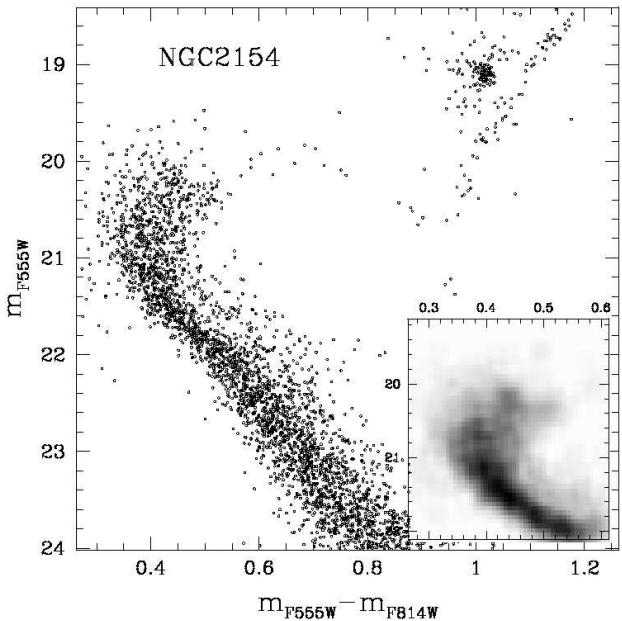}
      \caption{As in Fig.~\ref{NGC1846} for NGC~2154.}
         \label{ngc2154}
   \end{figure}
%__________________________________________________________________
%__________________________________________________________________
%
   \begin{figure}[ht!]
   \centering
   \includegraphics[width=8.75cm]{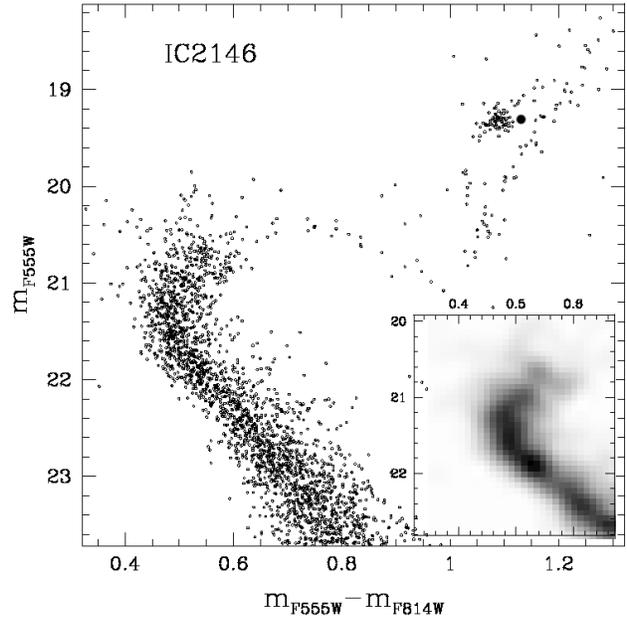}
      \caption{As in Fig.~\ref{NGC1846} for IC~2146.}
         \label{ic2146}
   \end{figure}
%__________________________________________________________________
%__________________________________________________________________
%
   \begin{figure}[ht!]
   \centering
   \includegraphics[width=8.75cm]{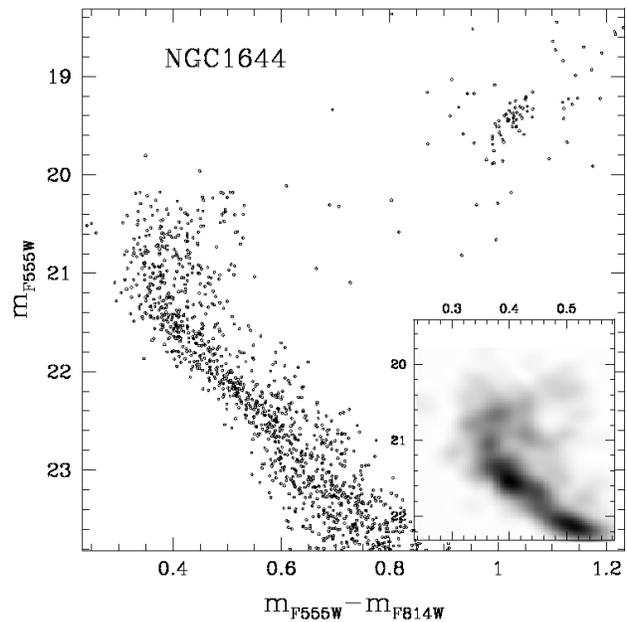}
      \caption{As in Fig.~\ref{NGC1846} for NGC~1644.}
         \label{ngc1644}
   \end{figure}
%__________________________________________________________________
%__________________________________________________________________
%
   \begin{figure}[ht!]
   \centering
   \includegraphics[width=8.75cm]{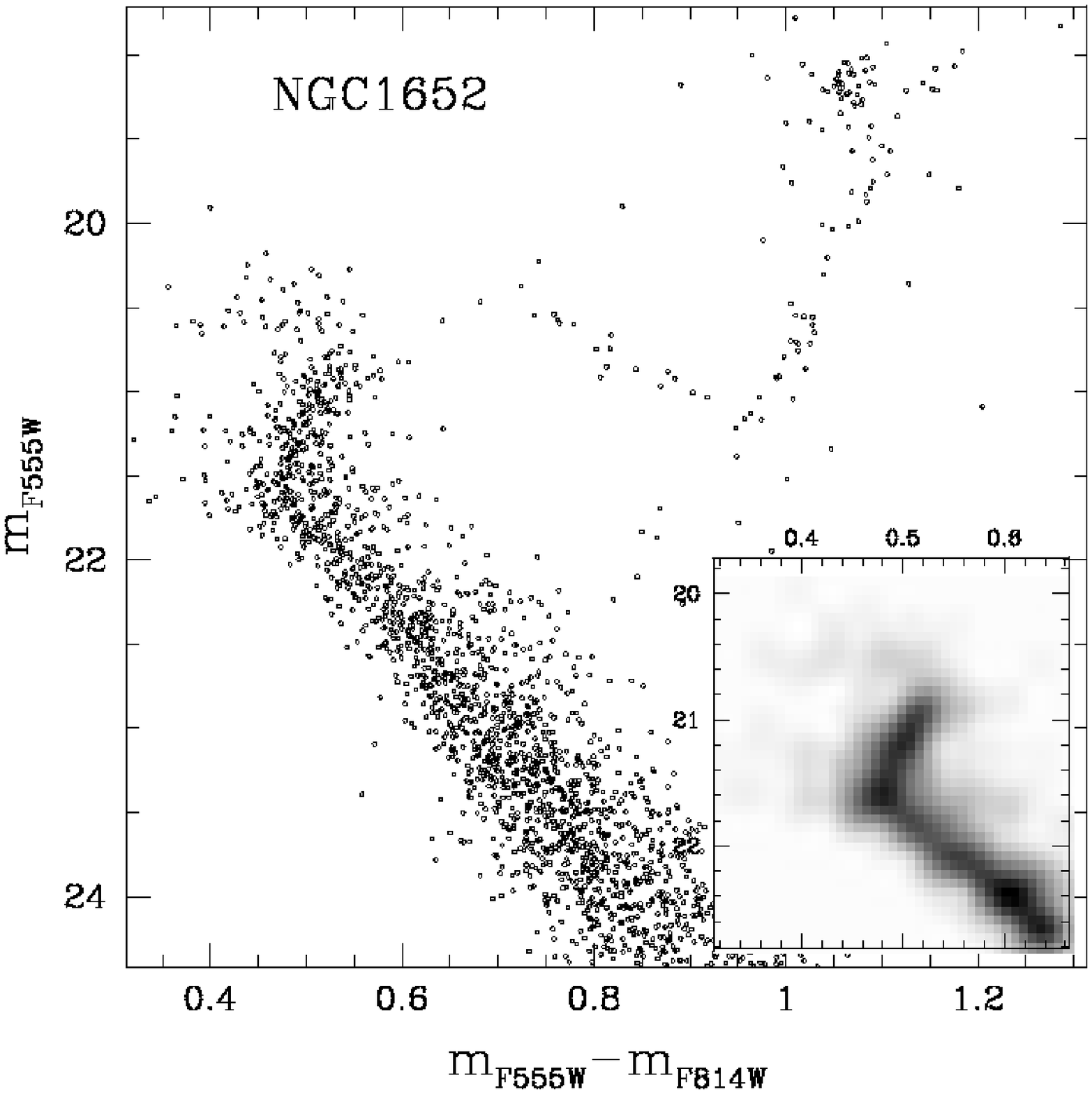}
      \caption{As in Fig.~\ref{NGC1846} for NGC~1652.}
         \label{ngc1652}
   \end{figure}
%__________________________________________________________________

\begin{figure}[ht!]
   \centering
   \includegraphics[width=8.75cm]{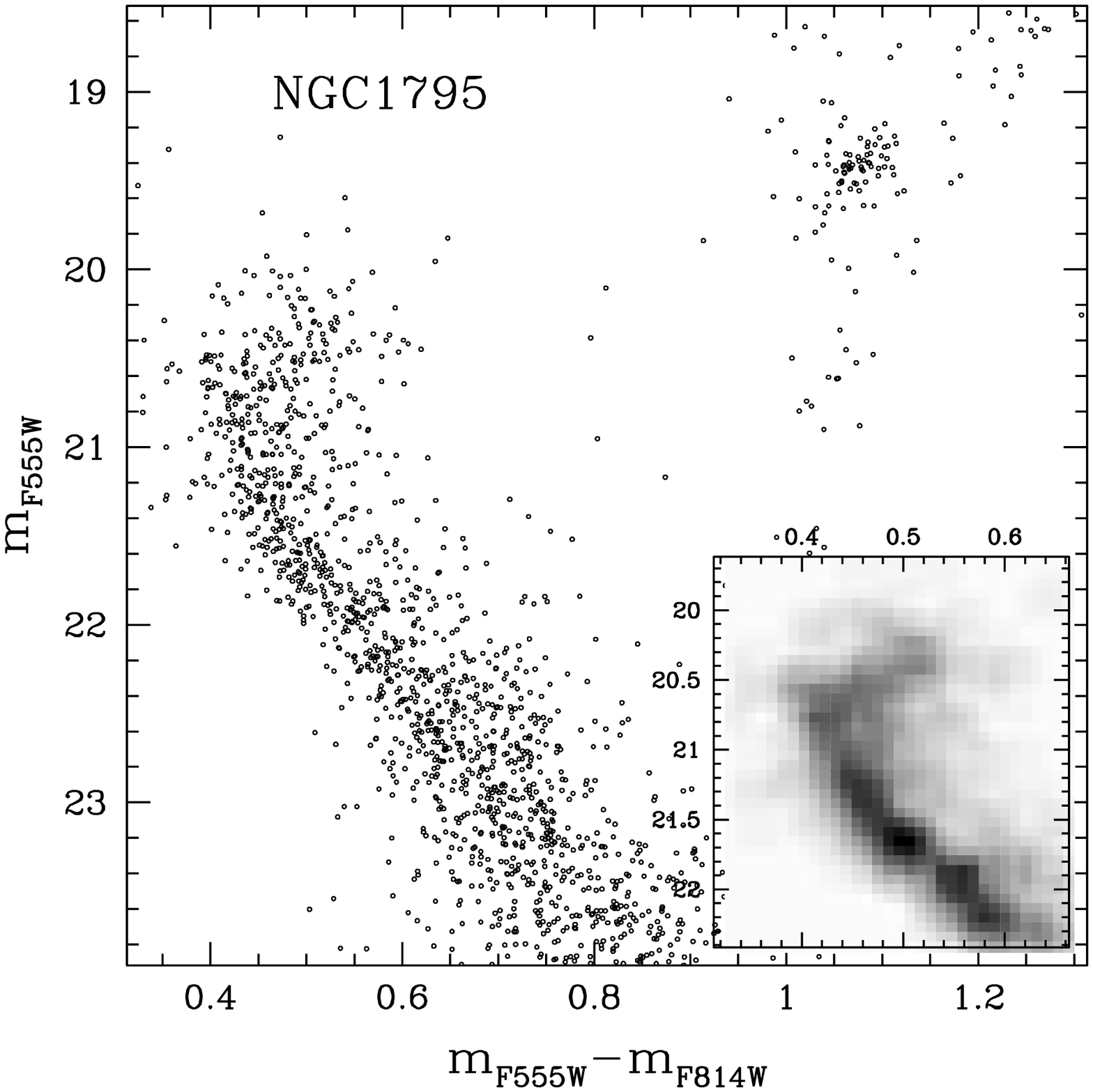}
      \caption{As in Fig.~\ref{NGC1846} for NGC~1795.}
         \label{ngc1795}
   \end{figure}

\begin{figure}[ht!]
   \centering
   \includegraphics[width=8.75cm]{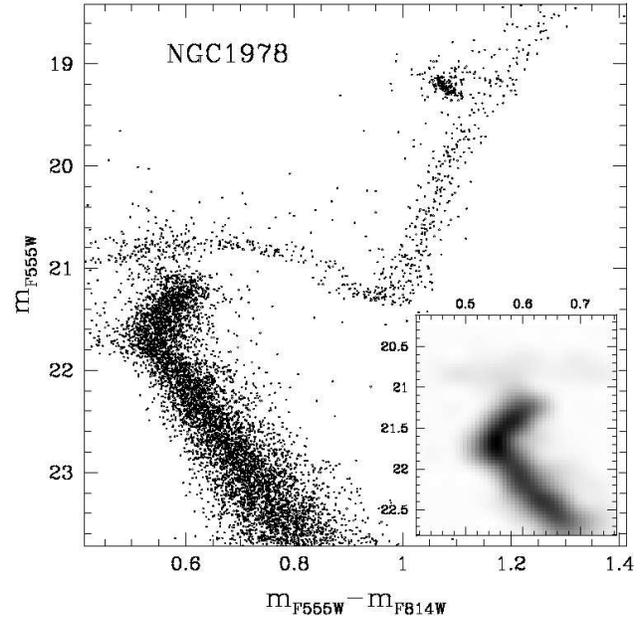}
      \caption{As in Fig.~\ref{NGC1846} for NGC~1978.}
         \label{ngc1978}
   \end{figure}

The   CMDs   around  the   MSTO   loci   for   the  remaining   twelve
intermediate-age  clusters that  have been  studied in  this  work are
presented in Fig.~\ref{eso} - \ref{ngc1978}.

Unfortunately, being this an archive project, the photometric data-set
available for these twelve clusters is not homogeneous.  One short and
two  deep images  for each  of the  three filters:  F435W,  F555W, and
F814W, were  collected within program GO~10595; while  one single deep
image  in  both  F555W  and  F814W bands  were  collected  within  the
snap-shot program  GO~9891.  Only two  clusters were observed  by both
programs  (NGC~1987  and  NGC~2108),  one  cluster  (ESO057-SC075)  in
GO-10595  only, and  the remaining  nine (HODGE~7,  IC~2146, NGC~1644,
NGC~1652,  NGC~1795, NGC~1852, NGC~1917,  NGC~1978 and  NGC~2154) were
observed within GO~9891 only (see Tab.~1 for more details).

We find that seven of these clusters show hints of an intrinsic spread
around the MSTO:  ESO057-SC075, HODGE~7, NGC~1852, NGC~1917, NGC~1987,
NGC~2108  and  NGC~2154.   The  CMDs of  IC~2146,  NGC~1644,  NGC~1652,
NGC~1795 and NGC~1978  show no evidence for such a  spread and are all
consistent  with  hosting  stars   with  the  same  age  and  chemical
composition, within our photometric precisions.
When simulated CMDs will be introduced, we will give a more objective
criterion to discriminate clusters with an intrinsic spread around the
MSTO (at the end of Section 6.2.2).

It must be noted that, apart from the broadened MSTO region, the other
primary features of  the CMDs of these clusters (MS,  RGB and AGB) are
narrow, and the  HB red-clump is well-defined, thus  the spread cannot
be  an artifact produced  by differential  reddening or  variations of
photometric zero points along the ACS field.

The limited statistic and photometric resolution do not allow us to
establish wheter, these spreads are just unresolved splits --as those
identified in the more populous clusters described in Sect.~4-- or not.

In the following  section, we show that the spread  around the MSTO of
the clusters mentioned  above must be intrinsic.  To  do this, we will
demonstrate  that the  broadening we  see cannot  be explained  by any
combination  of  photometric   errors,  field-star  contamination,  or
unresolved photometric binaries.
%__________________________________________________________________
%
%__________________________________________________________________
%
%__________________________________________________________________
%
%__________________________________________________________________

\section{Does the spreaded MSTO reflect the presence of multiple stellar populations? }
\label{sec:models}
Globular-cluster systems with multiple populations manifest themselves
in  many different  photometric ways.   In $\omega$  Cen  we (Anderson
1997, Bedin et  al.\ 2004) have detected a split of  the MS, which can
only be explained by two  stellar groups with different He content and
metallicity (Piotto  et al.\  2005), and also  at least  four distinct
SGBs  (which may  indicate  age  differences larger  than  1 Gyr,  see
Villanova  et  al.\  2007  and  references within).   In  NGC~2808  we
inferred the existence of  three distinct stellar populations from the
presence of three MSs (which  is most easily explained by three groups
of stars with different helium  content, see Piotto et al.\ 2007).  In
the  case of NGC~1851,  the presence  of two  populations of  stars is
inferred from  the fact that the  SGB splits into  two branches.  This
feature  can  be explained  either  by  two  distinct bursts  of  star
formation with  a time separation  of about 1  Gyr, or by  two stellar
populations  with distinct  initial  chemical composition  and a  much
smaller age difference (Milone et al.\ 2008, Cassisi et al.\ 2008).  A
splits of  the SGB have also  been observed in  NGC~6388 (Piotto 2008)
demonstrating that this massive  GC hosts two distinct stellar groups.
In a spectro-photometric study of M4, Marino et al.\ (2008) detected a
bimodal RGB, and demostrated that  it is due to a bimodal distribution
of  CN, Na,  O,  indicating that  also  this relatively-small  cluster
contains multiple populations.

The  Galactic clusters  cited  above variously  exhibit broadening  or
bifurcation of the RGB, SGB, and lower MS populations.  In the case of
the  intermediate-age  LMC clusters  we  have  studied  here, we  have
detected  both  splits  of  the  MSTO, as  in  NGC~1846  and  NGC~1806
(Sec.~\ref{sec:2MSTO}, M07 and M08), in NGC~1751 and possibly NGC~1783
(Sec.~\ref{sec:2MSTO}), and  a broadening of the MSTO  as the clusters
in  Figs.~\ref{eso}  -~\ref{ngc2108}  and  NGC~2173  (Bertelli  et  al
\ 2003).   The low numbers  of stars on  the RGB make it  difficult to
assess  the presence  of a  split or  anomalous broadening  along this
evolutionary sequence, and the distance of the LMC makes it impossible
to  detect  splits in  the  lower  MS  population with  the  presently
available data.

While  the  splits and  broadening  we  have  shown above  look  quite
convincing,  it   is  important  to  consider   the  possibility  that
photometric errors or other effects can generate anomalous spreads and
bifurcations, which  could be confused with differences  in age and/or
chemical composition.   In addition, it is also  important to consider
that both binaries  and field stars contaminate the  CMD region around
the  MSTO where  we are  most sensitive  to the  presence  of multiple
stellar populations.

In  Section~\ref{sec:fieldsubtraction}, we  demonstrate that  the MSTO
broadening  visible in  Figs.~\ref{eso}-~\ref{ngc2108} is  not  due to
field-star contamination by statistically subtracting field stars from
the cluster CMD.

In Sec.~\ref{sec:simCMD}, we consider  the influence that binary stars
could  have on  the  spread of  the MSTO.   To  do this,  we will  use
artificial stars  to simulate  the CMD of  a single population  plus a
population of binaries  such that visible in the  lower main sequence.
We then present the decontaminated cluster CMD and compare it with the
simulated one that includes both realistic errors and binaries to show
that  the observed  MSTO broadening  cannot be  explained by  a single
population.

\subsection{A method to decontaminate the cluster CMD from field stars}
\label{sec:fieldsubtraction}

We began our analysis looking at  the original CMD from stars near the
cluster  center, where we  estimated that  the contamination  by field
stars   is   at   most   at    the   level   of   $\sim$   20\%   (see
Section~\ref{sec:cmds}).

However,  it is possible  to reduce  this contamination  even  further by
doing a statistical  subtraction of the stars that  are most likely to
be field stars.  We have used the following approach which is based on
the assumption that the distribution of  field stars in the CMD of the
cluster and reference field is nearly  the same.  For each star in the
reference field  CMD, we  calculated the distance  to all  the cluster
field stars:  $d=\sqrt{(\Delta(m_{\rm F435W}-m_{\rm F814W}))^2+(\Delta
m_{\rm    F435W})^2}$   (or    $d=\sqrt{(\Delta(m_{\rm   F555W}-m_{\rm
F814W}))^2+(\Delta  m_{\rm  F555W})^2}$),   and  flagged  the  closest
cluster field star as a candidate to be subtracted.
We remind to the reader that 
cluster and reference fields cover the same area.

We determined the ratio $r$ between the completeness of this stars and
the completeness of the reference  field star. Note that this ratio is
always $r<1.0$.  In order  to avoid to  over-subtract field  stars, we
generated a random  number between 0 and 1, and  removed the star from
the cluster  CMD  if this random  number was less than  the ratio of
the completenesses.
%__________________________________________________________________
   \begin{figure}[ht!]
   \centering
   \includegraphics[width=8.75 cm]{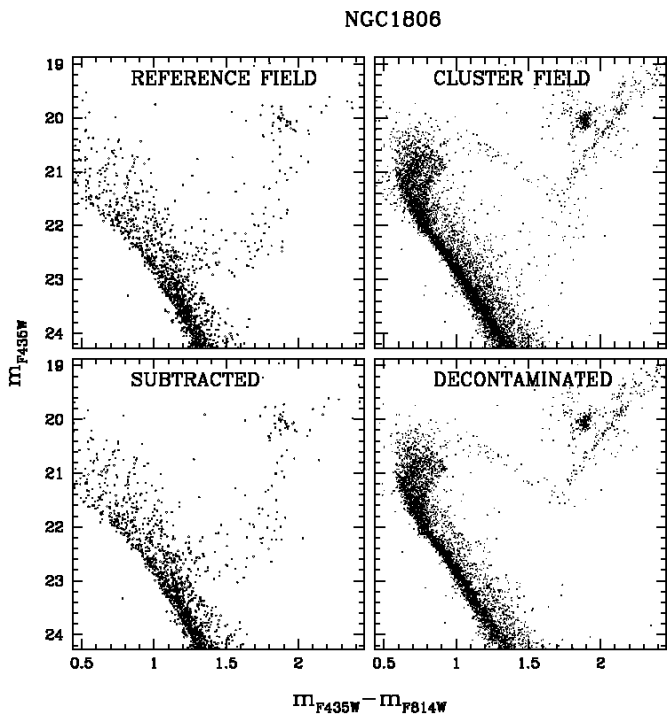}
      \caption{  Statistical decontamination from  field stars  of the
                CMD of NGC~1806. Upper  panels: CMD extracted from the
                reference (left), and  cluster fields (right).  Bottom
                panels: CMD of all the stars that have been subtracted
                from  the reference  field  (left) and  decontaminated
                cluster CMD (right).
      }
         \label{f3}
   \end{figure}
%__________________________________________________________________

The main steps of this  procedure are illustrated in Fig.~\ref{f3} for
NGC~1806:  The upper panels  show the  CMD of  stars in  the reference
(left)  and  cluster  (right)  field,  the  bottom  ones  the  CMD  of
subtracted  stars  (left) and  the  decontaminated  CMD (right).   The
statistical subtraction  of field  stars has not  been applied  to the
most populated clusters (NGC~1978, NGC~1783), which occupy most of the
field of view and have the highest ratio of cluster to reference field
stars.   We note  that there  is an  additional young  stellar cluster
within the  ACS images  of NGC~1852,  so we were  careful to  select a
reference field that was as far as possible from both clusters.

%__________________________________________________________________
   \begin{figure}[ht!]
   \centering
   \includegraphics[width=9 cm]{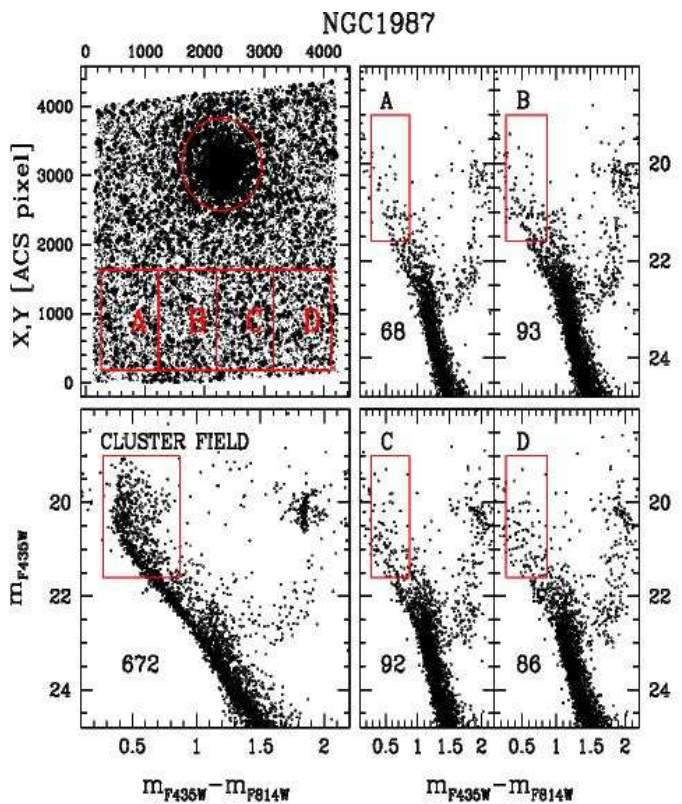}
      \caption{ Comparison of  the CMD of the cluster  field and those
        of four external reference regions ($A$, $B$, $C$ and $D$) for
        NGC~1987.  The values given in  the bottom left corner of each
        CMD are the number  of stars (corrected for incompleteness) in
        the box around the cluster MSTO.  }
         \label{4fields}
   \end{figure}
%__________________________________________________________________

Obviously,  small  variations in  the  distribution  of  stars in  the
reference field  are expected even  within the small WFC/ACS  field of
view. In order to check  whether the reference field star distribution
can be  reasonably assumed  uniform we defined  four regions  with the
same area of the cluster  field and the largest possible distance from
the cluster center.  We extracted  the CMD of each region and compared
it with that of the cluster field. A typical example of this procedure
is shown  in Fig.~\ref{4fields} for NGC~1987.  In  particular, we have
compared the number of stars (corrected for completeness) in the boxed
region around the MSTO.  We found that, in most cases the variation of
the  number  of stars  in  this region  is  smaller  than about  30\%.
Obviously, this procedure has been applied only to clusters that cover
a sufficiently  small fraction  of the ACS  field of view.   We cannot
exclude that the  variation of the number of  stars in reference field
regions  could   be  (partially)  due  to   contamination  of  cluster
members.  However, we  note that  oversubtracting the  cluster members
does not affect our results on the morphology of the MSTO.
%---------------------------------------------------------------

\subsection{The synthetic CMD}
\label{sec:simCMD}

The main  goal of  this section is  to generate  a synthetic CMD  of a
cluster with the same characteristics of the observed one, but hosting
a single stellar  population. This is a necessary  step to demonstrate
whether the observed  spread in color around the  MSTO is an intrinsic
feature  due to  the presence  of multiple  stellar populations  or an
artifact produced by unresolved binaries or photometric errors.

In  order to  simulate  a CMD  that  reproduces well  enough the  main
properties of the observed one we must ensure that:
\begin{itemize}
\item the mass function of the synthetic CMD is as similar as possible
  to the observed one;
\item simulated and observed  stars have the same radial distribution,
  completeness, and photometric errors;
\item  the  fraction and  the  mass  ratio  distribution of  simulated
  photometric binaries are as close as possible to the real ones.
\end{itemize}
In  the following  we  describe the  procedure  that we  have used  to
generate a synthetic  CMD with these requirements. It  consists of two
main steps: in  the first one we generate single  stars; in the second
one we  measure the fraction  of photometric binaries in  the observed
CMD and  add to the simulated  CMD the corresponding  number of binary
systems.

\subsubsection{Step one: simulation of single stars}
\label{simulation1}
We have separately simulated MS and evolved stars.\\

{\sc Unevolved Stars: }
To generate  single MS stars we  started by selecting  a sub-sample of
artificial stars  with the following  criterion. For each star  in the
cluster field that survive to the statistical subtraction of reference
field stars of  Section~\ref{sec:fieldsubtraction}, we have assigned a
sub-sample of the artificial stars  and randomly extracted a star from
it  (see Sect.~\ref{sec:artstars}  for a  description on  how  AS were
generated).   The  artificial-star  sub-sample  consists  of  all  the
artificial stars  with similar magnitudes (within  0.1 $m_{\rm F814W}$
mag) and  radial distances  (less then 50  pixel from  the observed
star).  This method produces a  catalog of simulated stars with almost
the same luminosity, and  radial distribution of the observed catalog.
This procedure has been applied only to MS stars.

{\sc  Evolved  Stars:  } 
Things become more complicated for  stars brighter than the MSTO where
the complex shape of the CMD  makes it harder to associate a star with
the CMD sequence that corresponds to its evolutionary phase.  For this
reason,  a  different approach  was  used  to  simulate evolved  stars
(i.e. stars brighter than the MSTO).

First  of  all, we  counted  the number  of  MS  stars (corrected  for
completeness)   with  $m_{\rm  F814W}^{\rm   TO}<m_{\rm  F814W}<m_{\rm
  F814W}^{\rm TO}+0.5$ and calculated  the average number of stars per
unit  mass $N_{\mathcal{M}}$  in this  range of  luminosity.  Next,  we obtained
$N_{\mathcal{M}}$ for  stars brighter than the  MSTO by using  a Salpeter (1955)
IMF and associated these stars to each portion of the CMD according to
the Pietrinferni et al.\ (2004) models.

Since evolved  stars should  all have nearly  the same mass,  we would
expect that  they would also  have the same radial  distribution.  For
this reason, we randomly associated to each simulated evolved star the
radial distance of an observed  star brighter than the MSTO.  Finally,
for each  simulated star, we  selected the sample of  artificial stars
with almost  the same magnitude,  color and distance from  the cluster
center (we imposed that both input $m_{\rm F814W}$ magnitude and color
must differ by less than 0.01  mag and radial distance by less than 50
pixels) and randomly extracted a star from it.

It must be noted that  the simulated CMDs have photometric errors that
are slightly smaller  than the errors of real  stars.  This reflects a
fundamental limitation of artificial-star tests: an artificial star is
measured by using the same PSF that was used to generate it, while for
the  real stars  we necessarely  have an  imperfect PSF.   The  PSF is
constructed to fit the real stars  as well as possible, but there will
invariably be errors  in the PSF model for the  real stars, which will
not be present for the artificial stars.  [However, it must be clearly
  stated  that these differences  are appreciable  only for  the stars
  with the highest S/N ratio].

Below, we describe the method  that we used to estimate the difference
between  the photometric errors  of real  and artificial  stars.  This
allows  us   to  introduce  an  additional  error   component  to  the
artificial-star photometry  so that the real  and artificial sequences
can be directly compared.  We illustrate the procedure for the case of
NGC~1806. The same procedure was applied for all clusters.

\begin{figure}[ht!]
   \centering
   \includegraphics[width=9 cm]{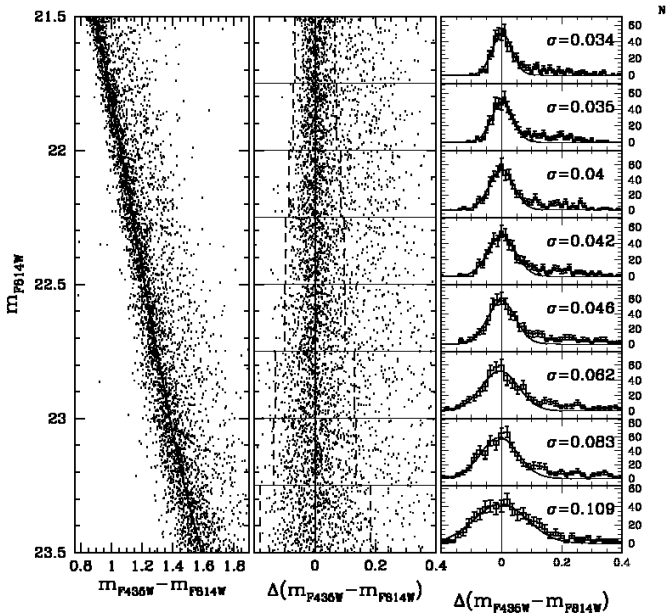}
      \caption{ $Left:$ CMD of NGC~1806 with the MSRL overplotted;
       $Middle:$ The CMD rectified by subtraction of the MSRL;
       $Right:$ Color distribution of the rectified CMD. The $\sigma$
       in the inset are those of the best-fitting gaussians.
       }
         \label{MScluster}
   \end{figure}

To determine how much  artificial broadening of the error distribution
is necessary,  we compare the  color distribution of the  observed and
simulated   MS   stars.    Figure~\ref{MScluster}  shows   the   color
distribution of MS stars for NGC~1806: the left panel contains the CMD
of all the stars contained  in the cluster field after the subtraction
of  the  reference field  stars.   The  red  fiducial line  (MSRL)  is
computed by using the following procedure.  We started by dividing the
CMD into  bins of magnitude  in the F814W  band and, for each  bin, we
calculated the median color and  magnitude and obtained a raw fiducial
line by fitting these points with  a spline.  Then we derived for each
star the  absolute value of the  difference between its  color and the
color  of  the fiducial  line,  and  calculated  the $\sigma$  as  the
$68.27^{\rm th}$ percentile. All  the stars with distances larger than
$N~\sigma$ from the  fiducial were rejected and the  survivors used to
redetermine the median color, magnitude and $\sigma$. We iterated this
procedure five times with $N$ going from 6 to 2 (integer numbers).

In  the  middle  panel  we  show  the  straightened  CMD  obtained  by
subtracting  from the color  of each  star the  color of  the fiducial
line.   In  the  right panel  we  show  the  histograms of  the  color
distribution  in  eight  $m_{\rm  F814W}$  magnitude  intervals.   The
distribution in color is well reproduced  by a Gaussian plus a tail on
the red side due to the conspicuous number of photometric binaries and
blends.

%__________________________________________________________________
%__________________________________________________________________
%__________________________________________________________________
   \begin{figure}[ht!]
   \centering
   \includegraphics[width=9 cm]{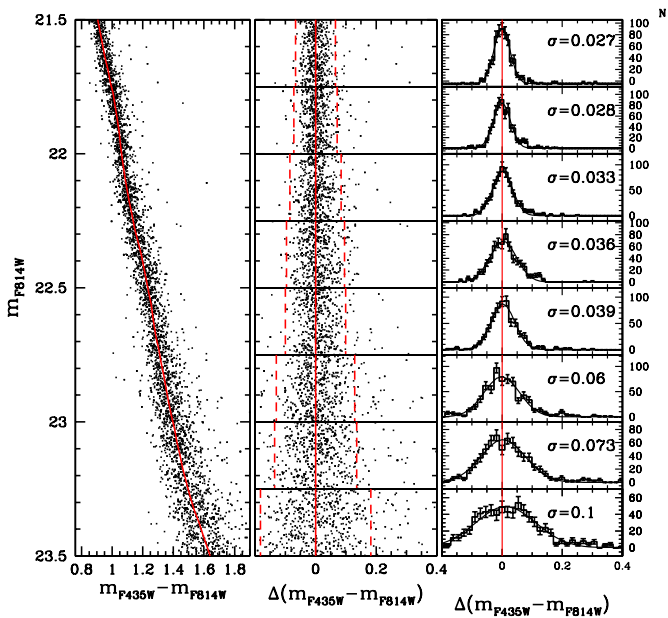}
      \caption{ As in Fig.~\ref{MScluster} but for artificial stars
       }
         \label{MSarts}
   \end{figure}
%__________________________________________________________________
%__________________________________________________________________
   \begin{figure}[!ht]
   \centering
   \includegraphics[width=8.75 cm]{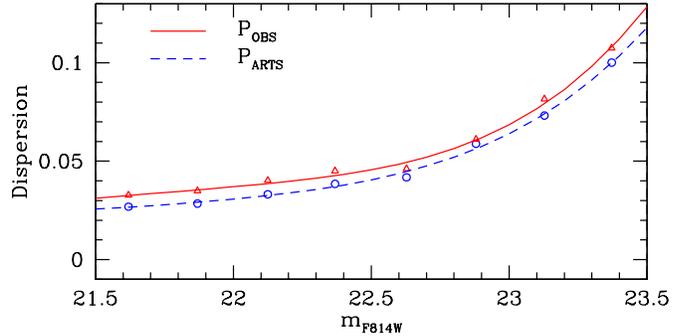}
      \caption{ Comparison of the  $\sigma$ of gaussians that best fit
        the  color distributions  of  MS stars  in  the observed  (red
        triangles)  and  artificial  stars  (blue circles)  CMD  as  a
        function  of  the $m_{\rm  F814W}$  magnitude. Continuous  and
        dashed lines are the best fitting fourth order polynomials.  }
         \label{curve}
   \end{figure}
%__________________________________________________________________

In Fig.~\ref{MSarts}  we also reproduce the distribution  in color for
the  artificial star MS.   As expected,  the spread  of the  latter is
slightly  lower  than that  of  the observed  MS.   We  note that  the
presence of this  additional dispersion, does not allow  us to exclude
intrinsic dispersion of the MSs smaller than ~0.03 mag in color, which
might result from dispersion of Z, Y, or a combination of the two (see
discussion in  Milone et al.\  2008).  In Fig.~\ref{curve}  we compare
the dispersions  of the observed MS ($\sigma_{\rm  OBS}$, circles) and
for artificial stars ($\sigma_{\rm  ART}$, triangles) as a function of
the  $m_{\rm F814W}$  magnitude;  the continuous  lines  are the  best
fitting  fourth order  polynomials ($P_{\rm  OBS}$ and  $P_{\rm ART}$).
Finally, we added to the color of each star of the artificial-star CMD
an  error  randomly  extracted   from  a  Gaussian  distribution  with
dispersion $P_{\rm DIF}=\sqrt{P_{\rm OBS}^{2}-P_{\rm ART}^{2}}$.

\subsubsection{Step two: simulation of binary systems}
\begin{figure}[!ht]
   \centering
   \includegraphics[width=9. cm]{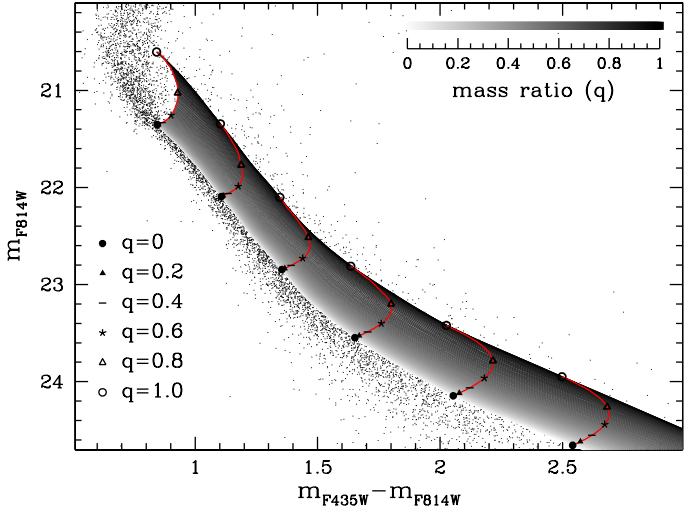}
      \caption{  Grey  area  highlight  the loci  populated  by  MS-MS
        binaries.  The levels  of grey  are proportional  to  the mass
        ratio. Red lines  indicate the position of a  MS-MS binary for
        six fixed values of the mass of the primary star ($m_{1}$) and
        q ranging from 0 to 1.  }
         \label{binTEO}
   \end{figure}

It is  clear from  the excess  of stars on  the red  side of  the main
sequence  that   many  of  these  clusters  have   a  sizeable  binary
populations (compare Fig.~26 with Fig.~27).   We expect  that this  binary population  can  have some
effect on  the distribution  of stars around  the turnoff, so  we will
estimate  the  binary-contamination   effect.   Binary  stars  of  LMC
clusters will  be unresolved  even with $HST$,  but the light  from each
star  will combine  and  the binary  system  will appear  as a  single
point-source object.  These  binaries can be discerned photometrically
from the single stars along the  MS as they are located brightward and
to the red of the sequence.  The position of the binary systems formed
by two  MS stars (MS-MS  binaries) with a  different mass ratios  in a
typical CMD  of an  intermediate age LMC  clusters are  illustrated in
Fig.~\ref{binTEO}.

%__________________________________________________________________
%__________________________________________________________________

In the  same region  of the  CMD, in addition  to the  true physically
associated  binaries,  we  expect  also a  few  chance  superpositions
(blends). The artificial-star tests (see  Fig.~27) show that this is a
very small effect.

It is important  to note here that measuring  the fraction of binaries
is  not the goal  of this  paper; our  aim here  is to  estimate their
fraction so that  we can determine how much of the  spread of the MSTO
region can reasonably be attributed to them.

\begin{figure}[ht!]
   \centering
   \includegraphics[width=9.5cm]{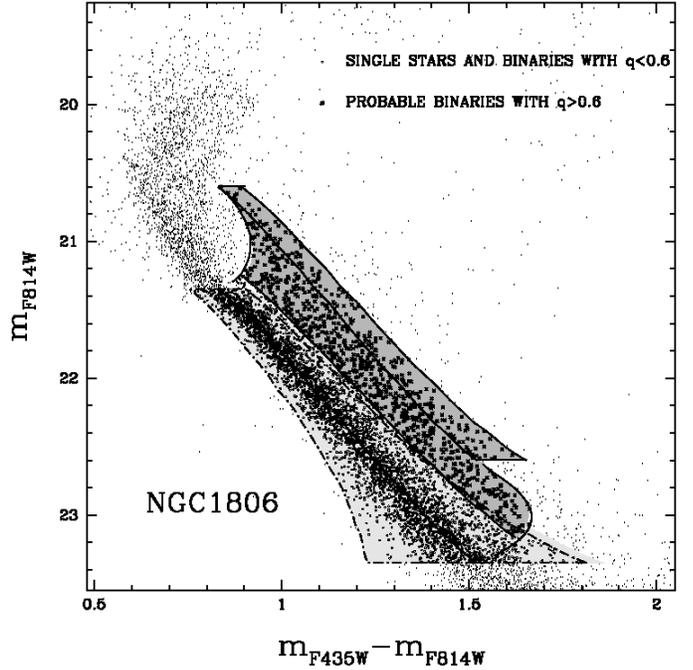}
      \caption{ The MS of NGC~1806 with the candidate binaries with
       mass ratio $q>0.6$ plotted as crosses.
       }
         \label{fig:binarie}
   \end{figure}

Binaries can be  parametrized by their mass ratio,  $q = m_{1}/m_{2}$,
where  $m_{1}<m_{2}$.  In  order  to estimate  the  fraction of  MS-MS
photometric binaries  with mass ratio  greater than a  threshold value
(hereafter $q_{\rm th}$,  see Tab.~2 for the adopted  $q_{\rm th}$ for
each  cluster) we  have applied  the  methods described  in Milone  et
al.\ (2008)  and Bedin et  al.\ (2008).  We  divided the CMD  into two
regions:  the first one  ($A$) contains  all the  single MS  stars and
binaries   with    a   primary   star    with   $(m_{\rm   F814W}^{\rm
  MSTO}+0.4)<m_{\rm  (F814W)}<(m_{\rm  F814W}^{\rm MSTO}+2.4)$  (where
$m_{\rm F814W}^{\rm MSTO}$  is the magnitude of the  MSTO in the F814W
band).   It   corresponds  to  the   light  and  dark  grey   area  of
Fig.~\ref{fig:binarie}.  The  second region ($B$)  is the part  of $A$
that contains  MS-MS binaries with $q>q_{\rm th}$,  and corresponds to
the  dark grey  area  of Fig.~\ref{fig:binarie}.   We adopted  $q_{\rm
  th}=0.6$ or  $q_{\rm th}=0.7$, as indicated in  Tab.~2, depending on
the photometric  quality of the  data.  The fraction of  binaries with
$q>q_{\rm th}$ has been evaluated as:
\begin{center}
$
f_{\rm BIN}^{q>q_{\rm th}}=\frac{N_{\rm CLUSTER}^{\rm B} - N_{\rm REFERENCE}^{\rm B}}
                        {N_{\rm CLUSTER}^{\rm A} - N_{\rm REFERENCE}^{\rm A} }
                 - \frac{N_{\rm ARTS}^{\rm B}} {N_{\rm ARTS}^{\rm A}}
$
\end{center}
%__________________________________________________________________
%__________________________________________________________________
Where $N_{\rm  CLUSTER}^{\rm A(B)}$ is the number  of stars (corrected
for completeness)  observed in the region  A (B) of  the CMD extracted
from  the  cluster  field.   $N_{\rm REFERENCE}^{\rm  A(B)}$  are  the
corresponding numbers of  stars in the CMD of  the reference field and
account for  the field contamination. $N_{\rm  ARTS}^{\rm A(B)}$ refer
to  artificial stars: their  ratio indicates  the fraction  of blends.
Finally,   we  calculated   the   global  fraction   of  binaries   by
interpolation,  assuming  a flat  mass-ratio  distribution.  The  only
differences  from what  done in  Milone et  al.\ (2008)  and  Bedin et
al.\ (2008)  is that  in this paper  we have removed  the contribution
from field  stars by using the  CMD observed in  the reference region,
rather than by using a Galactic model or proper motions.  The measured
fraction  of binaries  in each  cluster are  in Tab.~2.  

We tried very hard to get a reliable binary fraction for those
clusters, but unfortunately, their are just too big on ACS/WFC FOV.
If we take the outskirts as representative of the field we would
end up subtracting cluster members from clusters, and since energy
equipartion make the binary to sink into the cluster core, this
could potentially generate dangerous biases in the relative
fraction of single to binary system.  We avoid on purpose to give
numbers in Table 1 for those clusters, to avoid contamination of
the literature with unreliable values of the binary-fraction for
those objects.
%
%-----------------------------------
%
\begin{table}
%\centering
\label{tab_binarie}
\begin{tabular}{lccc}
\hline\hline  ID   & $q_{\rm th}$ & $f_{\rm bin}^{q>q_{\rm th}}$  & $f_{\rm bin}^{\rm TOT}$ \\
%------------------------------------------------------------------------
\hline
\hline
ESO057-SC075  & 0.6 & 0.17$\pm$0.02 & 0.42$\pm$0.05 \\
HODGE7        & 0.7 & 0.08$\pm$0.01 & 0.27$\pm$0.04 \\
%IC2146       &   &  &  &  \\
NGC1644       & 0.7 & 0.09$\pm$0.02 & 0.29$\pm$0.06 \\
NGC1652       & 0.7 & 0.06$\pm$0.01 & 0.19$\pm$0.04 \\
NGC1751       & 0.6 & 0.13$\pm$0.01 & 0.33$\pm$0.03 \\
%NGC1783      & & & \\
NGC1795       & 0.7 & 0.12$\pm$0.02 & 0.39$\pm$0.06 \\
NGC1806       & 0.6 & 0.13$\pm$0.01 & 0.32$\pm$0.03 \\
%NGC1846      & & &  \\
NGC1852       & 0.7 & 0.11$\pm$0.02 & 0.36$\pm$0.05 \\
NGC1917       & 0.7 & 0.09$\pm$0.02 & 0.31$\pm$0.05 \\
%NGC1978      &   &  &  &  \\
NGC1987       & 0.6 & 0.12$\pm$0.01 & 0.31$\pm$0.03 \\
NGC2108       & 0.6 & 0.18$\pm$0.01 & 0.46$\pm$0.03 \\
NGC2154       & 0.7 & 0.08$\pm$0.01 & 0.28$\pm$0.04 \\
%----------------------------------------------------------------------
\hline
\end{tabular}
\caption{Fraction of photometric binaries with $q>q_{\rm th}$ and total
  fraction of binaries (columns 5 and 6) within the cluster field. }
\end{table}
%----------------------------------------------------------------------

\begin{figure*}[ht!]
   \centering
   \includegraphics[width=\textwidth]{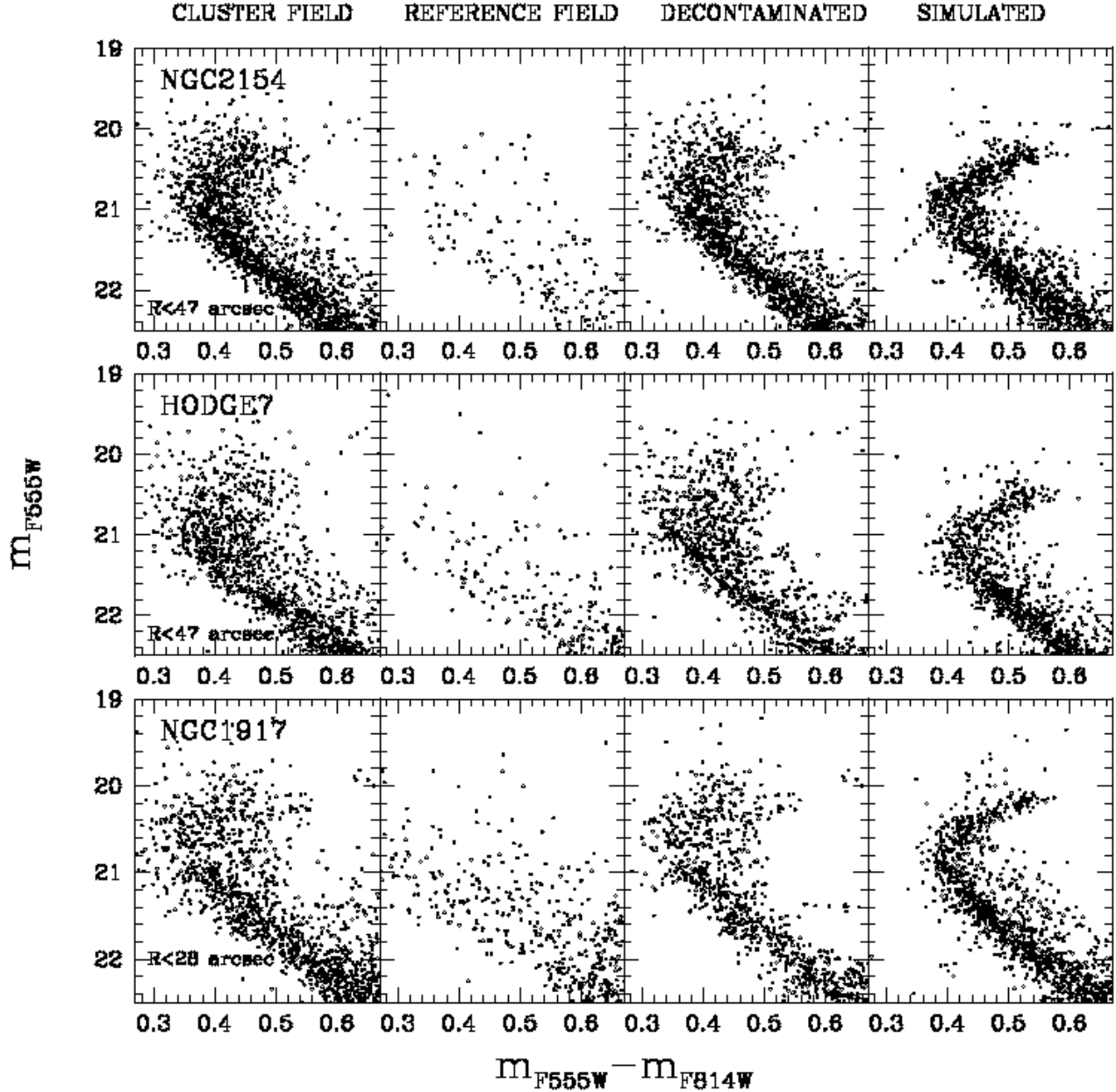}
      \caption{From the  left to the  right: CMD of the  cluster field
      for  NGC~2154,  HODGE~7 and NGC~1917,  CMD  of  the
      reference field,  CMD of the cluster field  after that reference
      field stars  have been statistically  subtracted, simulated CMD.
      }
         \label{compA}
   \end{figure*}

\begin{figure*}[ht!]
   \centering
   \includegraphics[width=\textwidth]{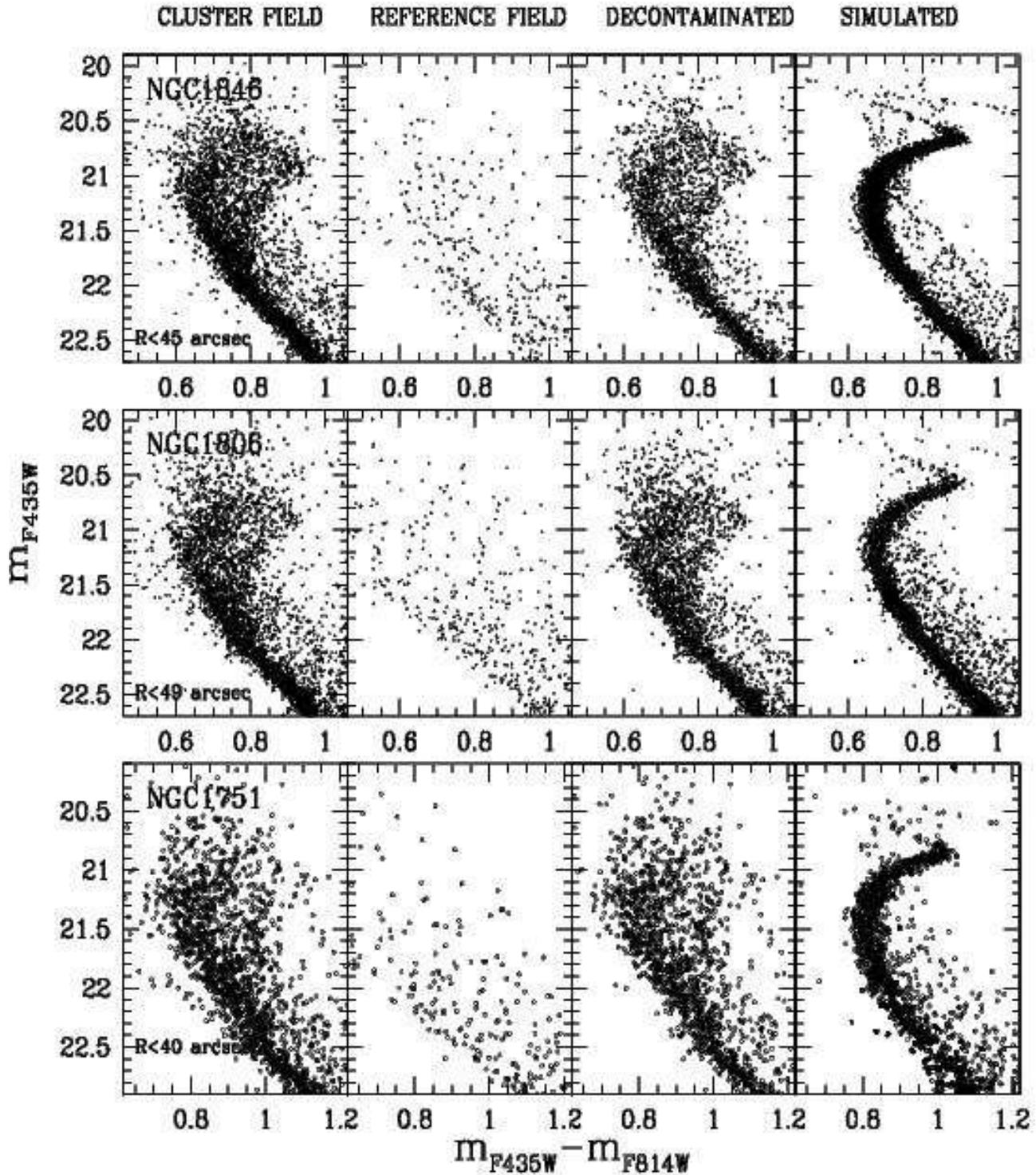}
      \caption{As in Fig.~\ref{compA} for NGC~1846, NGC~1806 and NGC~1751. }
         \label{compB}
   \end{figure*}

\begin{figure*}[ht!]
   \centering
   \includegraphics[width=\textwidth]{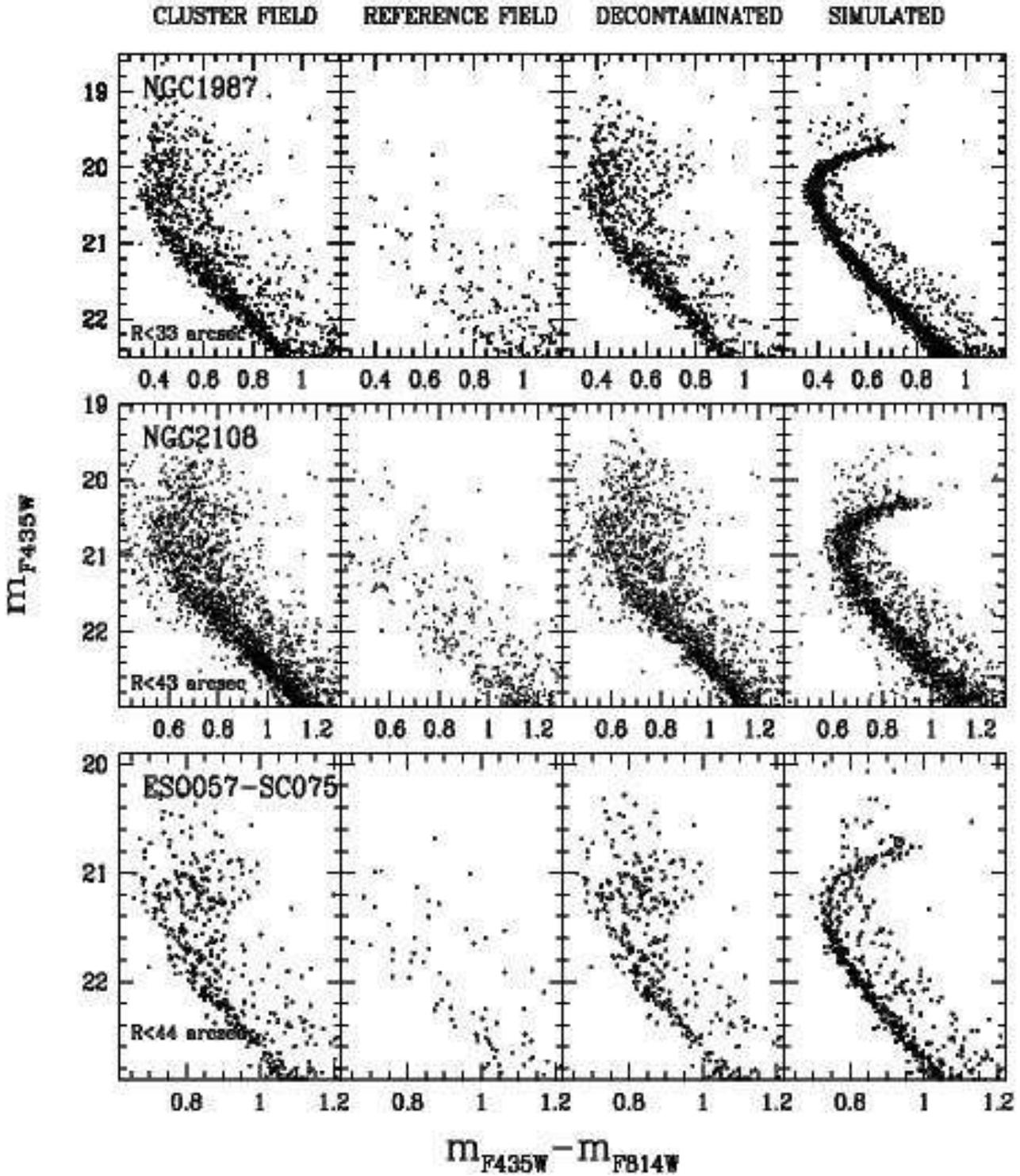}
      \caption{As in Fig.~\ref{compA} for NGC~1987, NGC~2108 and
      ESO057-SC075. }
         \label{compC}
   \end{figure*}

\begin{figure*}[ht!]
   \centering
   \includegraphics[width=\textwidth]{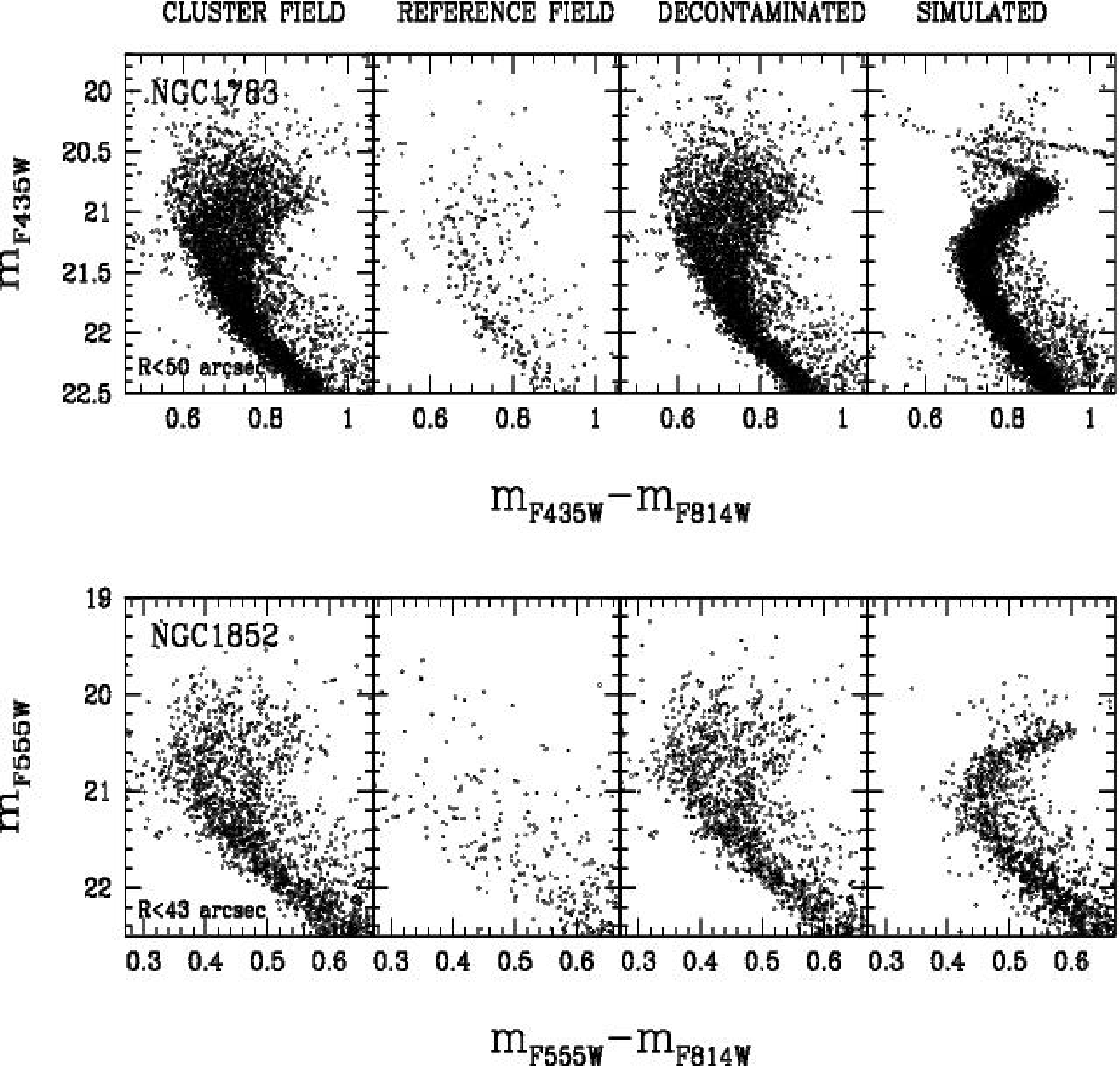}
      \caption{As in Fig.~\ref{compA} for NGC~1783 and NGC~1852. }
         \label{compD}
   \end{figure*}

To simulate binary stars to be added to the simulated CMD described in
Sec.~\ref{sec:simCMD} we adopted the following procedure:
\begin {itemize}
\item{We     selected    a     fraction     $F_{\rm    BIN}$     
of  single   stars  equal  to  the  measured
  fraction  of  binaries  and   derived  their  masses  by  using  the
  Pietrinferni  et  al.   (2004)  mass-luminosity  relation  (for  the
  clusters where the  measure of the binary fraction  is not available
  we assumed the average value of $F_{\rm BIN}=0.33$);}
\item{For each of them, we calculated the mass $\mathcal{M}_{2}=q \times \mathcal{M}_{1}$ of the
  secondary star and  obtained the corresponding  $m_{\rm F814W}$
  magnitude. Its  color was derived by the MSRL;}
\item {Finally, we summed up the F435W (or F555W) and F814W fluxes of the two
  components, calculated the corresponding magnitudes, added the
  corresponding photometric error, and replaced in the CMD the original
  star with this binary system.}
\end{itemize}
Figures~\ref{compA}, ~\ref{compB},  ~\ref{compC} and ~\ref{compD} show
the contribution that we would  expect from binaries to the broadening
of the  cluster TOs.  It  is clear that  the double MSTO  in NGC~1806,
NGC~1846,  and   NGC~1751,  and  the  extended   (broadened)  MSTO  in
ESO057-SC075,   HODGE~7,  NGC~1783,   NGC~1852,   NGC~1917,  NGC~1987,
NGC~2108  and NGC~2154 are  intrinsic features  of these  objects, and
cannot  due   to  field-star  contamination,   photometric  errors  or
binaries.  In  these figures, we show,  from the left to  the right, a
zoom  of the  original cluster-field  CMD  around the  MSTO, the  same
portion of the CMD but for stars in the reference field, the CMD after
field-star decontamination, and the  simulated CMD for these clusters.

Figure~35 and 36 illustrate the same exercise for the clusters with no
significant MSTO broadening.
Now that we have introduced the simulated  CMDs we can describe how we
discriminate between clusters that show evidence of hosting a multiple
population,  and  those which  do    not.  First,  we  calculated  the
dispersion   of  stars along a  same    direction perpendicular to the
obseverd     spread (on     the  right  of    the    MSTO) for    real
($\sigma_{SGB}^{OBS}$) and artificial  stars   ($\sigma_{SGB}^{ART}$).
Then, we considered a cluster as  hosting a multiple population if the
dispersion of  real stars was more than  three times the dispersion of
the  simulated CMD.  This  condition is  verified for  eleven  out the
sixteen clusters.  Although the exact position of the line along which
to measure the spread   is not unique,   it seemed to us  a reasonably
solid approach.

%--------------------------
%__________________________________________________________________
%
%__________________________________________________________________
%__________________________________________________________________
%
%__________________________________________________________________
%__________________________________________________________________
%
%__________________________________________________________________
%__________________________________________________________________
%
%__________________________________________________________________

\section{Isochrones fitting }
\label{sec:iso}
M07  and M08  proposed that  the  split of  the MSTO  of NGC~1846  and
NGC~1806  is consistent  with  the presence  of  two distinct  stellar
populations with the same chemical composition and a difference in age
of about $\sim300$ $\rm Myr$. In addition, M08 have suggested that the
spread of the  MSTO of NGC~1783 can be attributed  to a prolonged star
formation.

\begin{figure*}[ht!]
   \centering
   \includegraphics[width=\textwidth]{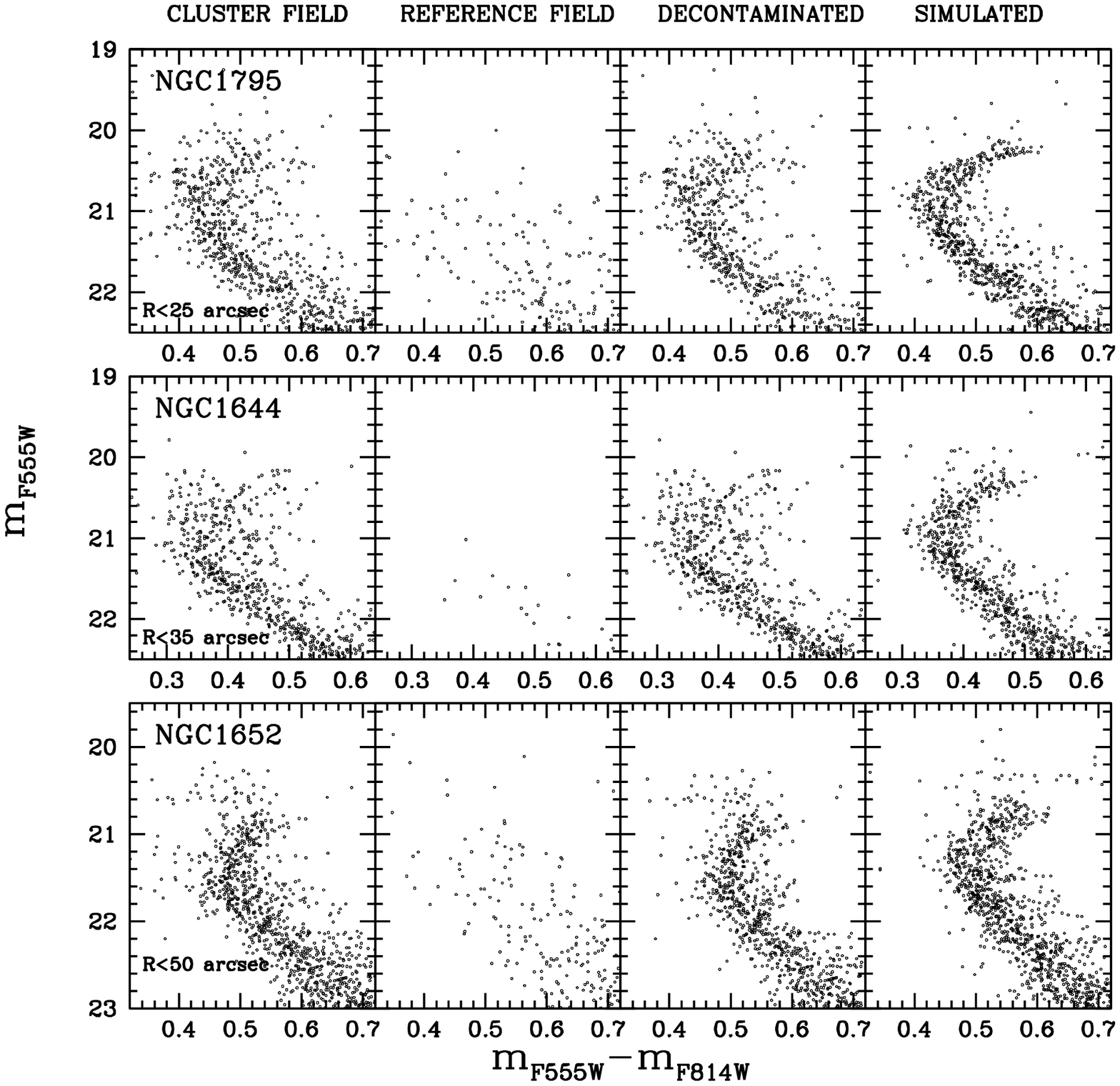}
      \caption{As in Fig.~\ref{compA} for
      NGC~1975, NGC~1644 and NGC~1652. }
         \label{compE}
   \end{figure*}

\begin{figure*}[ht!]
   \centering
   \includegraphics[width=\textwidth]{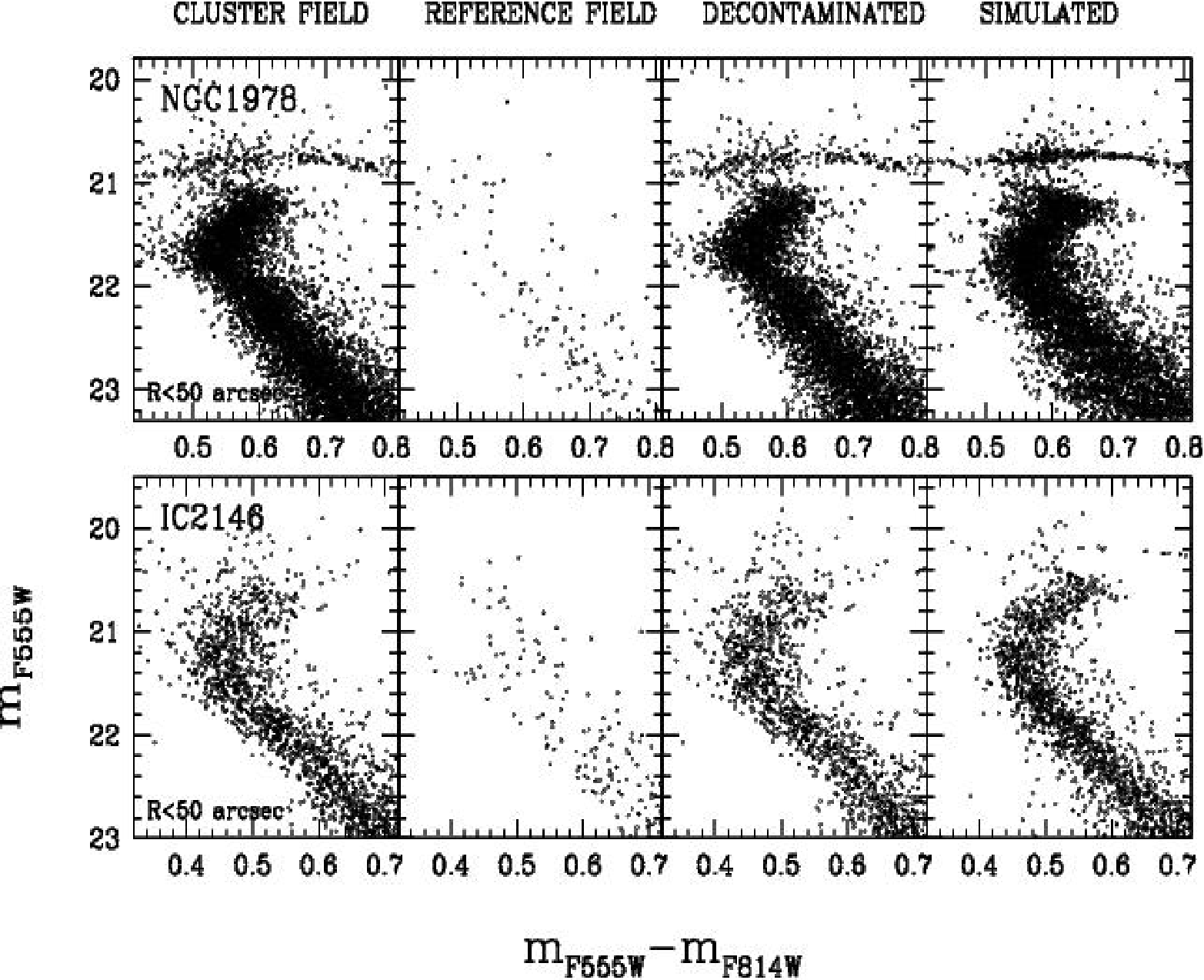}
      \caption{As in Fig.~\ref{compA} for NGC~1978 and IC~2146. }
         \label{compF}
   \end{figure*}

In the  previous sections, we have demonstrated  that eight additional
clusters show  strong indications of  an intrinsic spread or  split of
the  CMD around the  MSTO region.  The remaining  five objects  of our
sample show  no significant  evidence for multiple  stellar population
within   our   photometric   precision   (see   Fig.~\ref{compE}   and
~\ref{compF}).
%__________________________________________________________________
%
%__________________________________________________________________
%__________________________________________________________________
%
%__________________________________________________________________
%

We note  that in the clusters where  a spread in the  MSTO is claimed,
the other evolutionary sequences are  narrow and well defined, after a
correction for differential reddening has been made.  The tightness of
these other sequences  is also a strong indication  that there is very
little variation in metallicity among the stars.

Therefore,  in  the  absence  of  any  detailed  chemical  composition
analysis, in  what follows we will  assume that any  spread around the
MSTO can  be attributed to a  difference in age alone.  We will derive
the  main  parameters for  each  cluster  (metallicity,  age, and  the
maximum age spread among the populations) by fitting the data with the
isochrones  from the  BaSTI  evolutionary code  (Pietrinferni et  al.\
2004, updated version  of August 2008).  Each of  the isochrones has a
solar-scaled   distribution   of   metals  and   includes   convective
overshooting.

To determine the  isochrone that best matches the  observed CMD of our
clusters with  no evidence of an  intrinsic spread around  the MSTO we
followed a procedure similar to  the one adopted by M07.  We generated
a  grid of  isochrones  using  metallicities of  Z=  0.008 and  0.010,
sampling an  age range between 1.0  and 3.0 Gyrs at  intervals of 0.05
Gyr.  Then, we defined by hand:\  the magnitude of the MSTO, the color
of the RGB at a level intermediate  between that of the red end of the
SGB  and  that of  the  HB  red-clump, and  the  magnitude  of the  HB
red-clump.   In this  way we  calculated the  difference  in magnitude
between the  MSTO and  the HB red-clump  ($\Delta_{\rm mag}$)  and the
difference in  color between the MSTO  and the fiducial  points on the
RGB ($\Delta_{\rm  col}$). Then we calculated a  value of $\Delta_{\rm
  mag}$  and $\Delta_{\rm  col}$ for  each isochrone  in the  grid and
compared them  with the observed  ones.  Finally, we selected  all the
isochrones  where  $\Delta_{\rm col}$  and  $\Delta_{\rm mag}$  differ
respectively  by less  $\pm$0.25 and  $\pm$0.03 magnitudes  and fitted
them to the  CMD by hand.  To do this, we  varied the distance modulus
in the  range $18.30<(m-M)_{0}<18.70$  and the reddening  between 0.00
and 0.30, both  in steps of 0.01 mag and  searched for the combination
that best matches the cluster sequences.
 For  clusters with  a double  or broadened  MSTO, we  used  a similar
 approach with the  exception that, in this case,  we first defined on
 the  observed  CMD  the magnitude  of  the  MSTO  and the  values  of
 $\Delta_{\rm  mag}$   and  $\Delta_{\rm  col}$  for   the  bMSTO  and
 determined  the isochrone  that  best fits  this younger  population.
 Then,  we selected  all  the isochrones  with  the same  metallicity,
 distance  modulus and reddening,  but different  ages and  fitted the
 fMSTO (for clusters with a  broadened MSTO we calculated those values
 that correspond to the brighter and the fainter region of the MSTO).

The best fitting distance  modulus, reddening, metallicity and age are
listed  in  Tab.~3.   The   last  column  indicates  the  maximum  age
difference,  $\Delta_{\rm{age}}$,  for  stars  in clusters  that  show
possible evidence of multiple or prolonged star formation episodes.

\begin{table}[ht!]
%\centering
\label{t1}
\scriptsize{
\begin{tabular}{lcccll}
\hline\hline  ID  &  $(m-M)_{0}$ & $\it{E(B-V)}$ & Z  &  Age (Myr) & $\Delta_{\rm{age}}$ \\
%------------------------------------------------------------------------
\hline
\hline
ESO057-SC075 & 18.46 & 0.14 & 0.008 & 1400-1600 & $200\pm50$ \\
HODGE7  & 18.48 & 0.04 & 0.008 & 1400-1550 & $150\pm50$ \\
IC2146  & 18.50 & 0.07 & 0.008 & 1550      & $<50$      \\ 
NGC1644 & 18.48 & 0.01 & 0.008 & 1550      & $<50$      \\ 
NGC1652 & 18.48 & 0.06 & 0.008 & 1700      & $<50$      \\ 
NGC1751 & 18.45 & 0.22 & 0.008 & 1300-1500 & $200\pm50$ \\
NGC1783 & 18.46 & 0.06 & 0.008 & 1400-1600 & $200\pm50$ \\
NGC1795 & 18.45 & 0.10 & 0.008 & 1300      & $<50$      \\ 
NGC1806 & 18.44 & 0.09 & 0.008 & 1400-1600 & $200\pm50$ \\
NGC1846 & 18.49 & 0.09 & 0.008 & 1350-1600 & $250\pm50$ \\
NGC1852 & 18.50 & 0.08 & 0.008 & 1200-1450 & $250\pm50$ \\
NGC1917 & 18.48 & 0.08 & 0.008 & 1200-1350 & $150\pm50$ \\
NGC1978 & 18.49 & 0.09 & 0.008 & 2000      & $<100$     \\ 
NGC1987 & 18.40 & 0.04 & 0.010 &  950-1200 & $250\pm50$ \\
NGC2108 & 18.40 & 0.21 & 0.010 &  950-1100 & $150\pm50$ \\
NGC2154 & 18.48 & 0.04 & 0.008 & 1350-1500 & $150\pm50$ \\
%----------------------------------------------------------------------
\hline
\end{tabular}
\caption{Parameters that have been used to obtain the best fit between
the observed CMD and the BaSTI isochrones.}
}
\end{table}
%----------------------------------------------------------------------

\begin{figure*}[ht!]
   \centering
   \includegraphics[width=\textwidth]{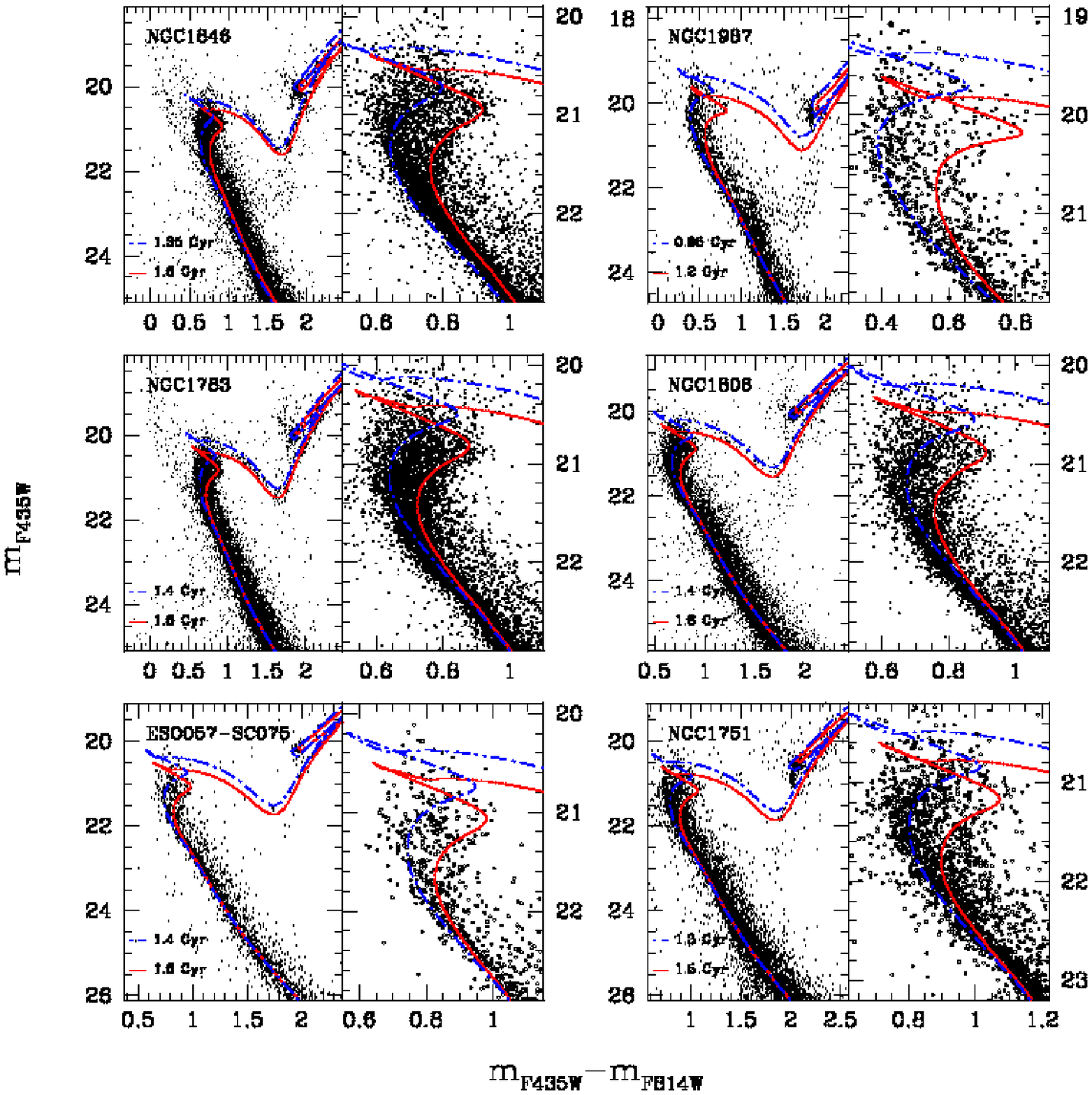}
      \caption{The  best  fitting isochrones,  obtained  by using  the
        distance  modulus,   reddening,  metallicity  and   age(s)  of
        NGC~1846,  NGC~1987,   NGC~1783,  NGC~1806,  ESO057-SC075  and
        NGC~1751 listed  in Table~\ref{t1} are overplotted  to the CMD
        of the  cluster field of (left).  A zoom of  the region around
        the MSTO is shown on the right.}
         \label{iso1}
   \end{figure*}

\begin{figure*}[ht!]
   \centering
   \includegraphics[width=\textwidth]{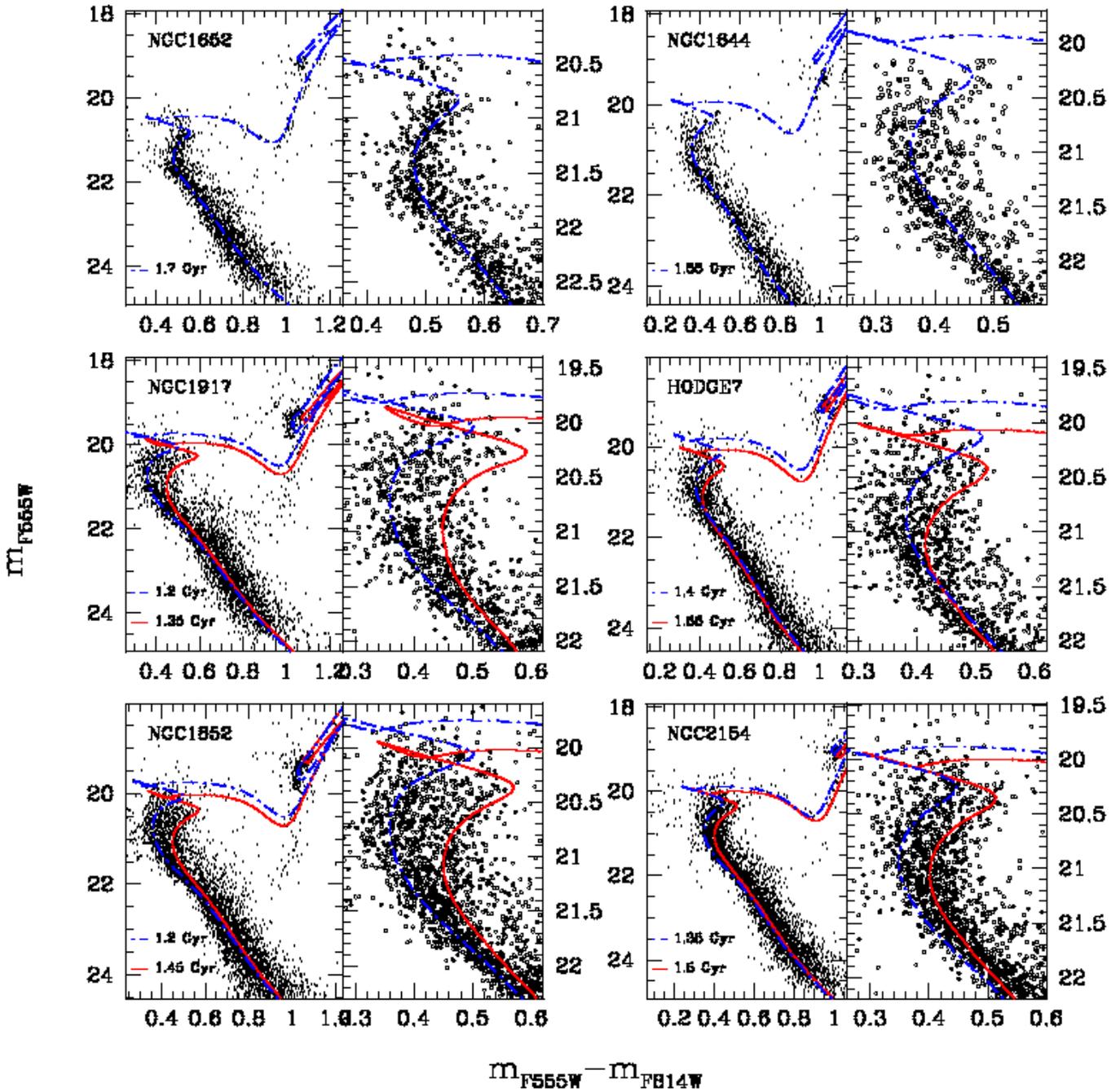}
      \caption{As in Fig.~\ref{iso1} for NGC~1652, NGC~1644, NGC~1917,
      HODGE7, NGC~1852 and NGC~2154.}
         \label{iso2}
   \end{figure*}

\begin{figure*}[ht!]
   \centering
   \includegraphics[width=\textwidth]{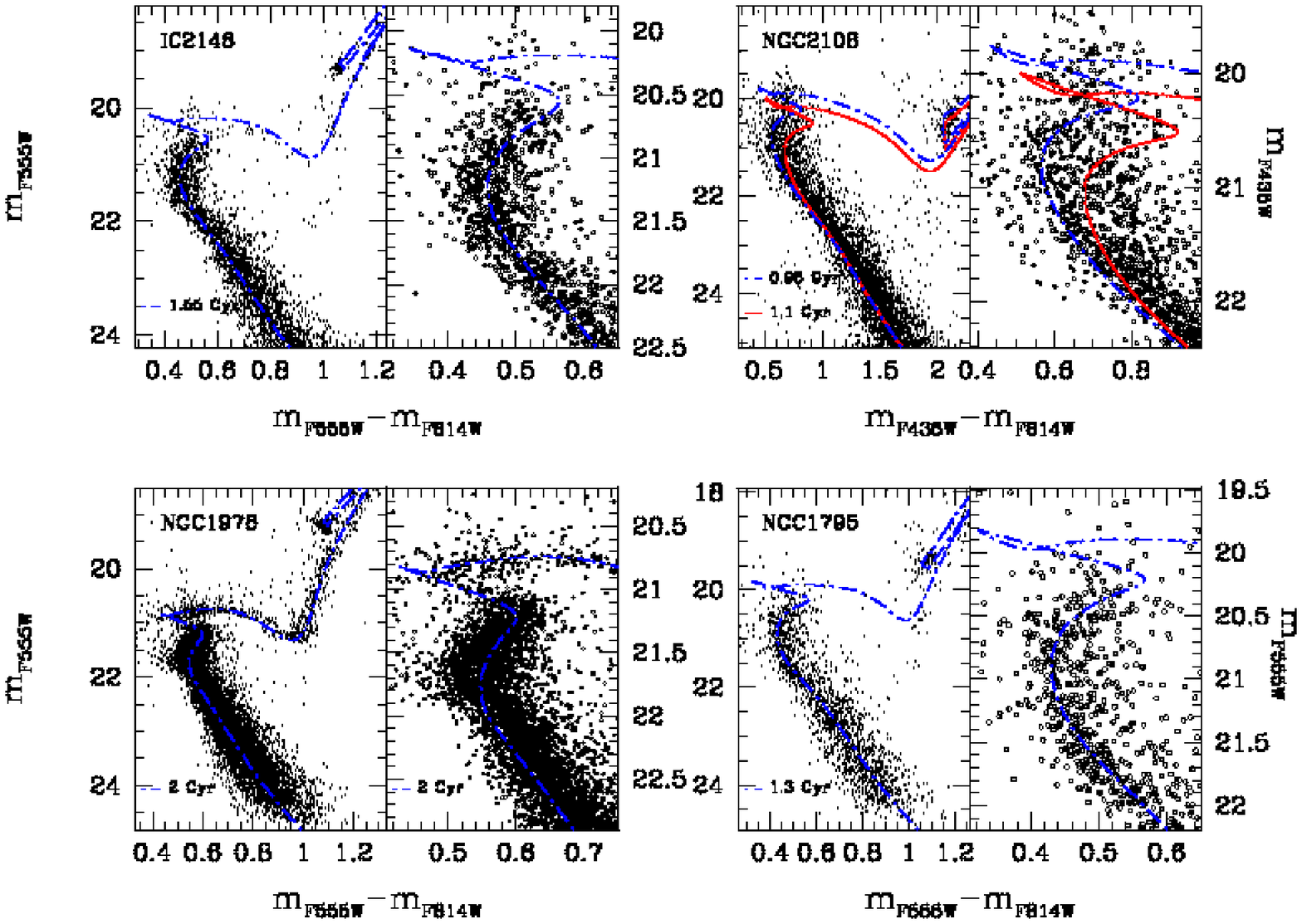}
      \caption{As in Fig.~\ref{iso1} for IC~2146, NGC~2108, NGC~1978
      and NGC~1795. }
         \label{iso3}
    \end{figure*}

In  Fig.~\ref{iso1},  Fig.~\ref{iso2}  and  Fig.~\ref{iso3},  we  have
overplotted  to  the  observed   CMDs  the  best  fitting  isochrones.
Interestingly,  the multiple  (or  prolonged) star-formation  episodes
seem to lay between 150 and 250 Myr, very similar to the time interval
between successive  star formation  episodes in the  intermediate mass
AGB star ejecta pollution depicted by Ventura et al (2001).

%__________________________________________________________________
%
%__________________________________________________________________
%__________________________________________________________________
%
%__________________________________________________________________
%__________________________________________________________________
%
%__________________________________________________________________
\section{Conclusions}
\label{sec:conclusion}
High  precision $HST$ ACS/$HST$  photometry  of sixteen  intermediate age  LMC
stellar clusters has revealed that eleven of them (i.e. about the 70$\pm$25\%
of the entire sample) host multiple stellar populations.

The  CMDs of  NGC~1806, NGC~1846  and NGC~1751  exhibit  two dinstinct
MSTOs, suggesting that these  clusters have experienced (at least) two
main episodes  of star formation with  a temporal separation  of 200 -
250 Myr.  For these three  clusters the high quality of our photometry
enabled us, not  only to distinguish the two  populations, but also to
measure the fraction  of stars belonging to each of  them.  In all the
cases, the population corresponding  to the brighter MSTO (the younger
population) is  the main population, and includes  more then two-third
of cluster  stellar population, consistent with  the intermediate mass
AGB pollution scenario (See D'Antona \& Caloi 2008 for a recent review).

Our photometry strongly suggests  that the intrinsic broadening of the
MSTO of NGC~1783  observed by M08 could be  attributed to the presence
of two distinct branches which  are closely spaced and poorly resolved
by the observations.

In seven  additional clusters, namely  ESO057-SC075, HODGE7, NGC~1852,
NGC~1917, NGC~1987,  NGC~2108, and NGC~2154 we observed  a wide spread
in color  for the stars around the  MSTO. In spite of  this, the other
main features  of the CMD  are narrow and  well-defined, demonstrating
that  the  spread  cannot  be  an artifact  produced  by  differential
reddening, by variation of the  photometric zero point along the chip,
or by a relatively large spread  in metallicity.  By using the CMD of  the stars in the
fields  that surround the  cluster, we  demonstrate that  the observed
feature  is  unequivocally  associated  with the  clusters.   Finally,
artificial stars and simulated CMDs show that the wide spread in color
observed around the  MSTO cannot be produced by  photometric errors or
binaries.  It is interesting to  note that the age spreads observed in
this sample appear  to be quite similar: they are  all between 150 and
250 Myr.

%_____________________________________________________________________
%
\begin{acknowledgements}
We thank the anonymous referee for the careful reading of the manuscript,
and for the useful comments.
\end{acknowledgements}
%_____________________________________________________________________
%

%%%%%%%%%%%%%%%%%%%%%%%%%%
%
\bibliographystyle{aa}

\begin{thebibliography}{}
\bibitem[()]{Anderson97} Anderson, J. 1997, Ph.D. thesis,
  Univ. California, Berkeley
\bibitem[()]{Anderson06} Anderson, J. \& King, I. \ R., 2006. ACS/ISR
  2006-01, PSFs, Photometry and Astrometry for the ACS/WFC
\bibitem[()]{Anderson08} Anderson, J.\ et al. 2008, AJ, 135, 2055
\bibitem[()]{Bedin04} Bedin, L. \ R., Piotto, G., Anderson, J.,
  Cassisi, S., King, I. \ R., Momany, Y., \& Carraro, G., 2004, \apj,
  605, L125
\bibitem[()]{Bedin05} Bedin, L. \ R., Cassisi, S., Castelli, F.,
  Piotto, G., Anderson, J., Salaris, M., Momany, Y. \& Pietrinferni,
  A., 2005, MNRAS, 357, 1048
\bibitem[()]{Bedin08} Bedin, L. \ R., Salaris, M., Piotto, G.,
  Cassisi, S., Milone, A. \ P., Anderson, J., \& King, I. \ R., 2008,
  \apj, 679, L29
\bibitem[()]{Bertelli03} Bertelli, G. Nasi, E., Girardi, L., Chiosi,
  C., Zoccali, M., \& Gallart, C. 2003, AJ, 125, 770
\bibitem[()]{Cassisi08} Cassisi, S., Salaris, M., Pietrinferni, A.,
  Piotto, G., Milone, A. \ P., Bedin, L. \ R., \& Anderson, J., 2008, \apj,
  672, L115
\bibitem[()]{DAntona05} D'Antona, F., Bellazini, M., Caloi, V., Fusi
  Pecci, F., Galleti. S., \& Rood, R. \ T., 2005, \apj, 631, 868
\bibitem[()]{DAntona08} D'Antona, F., \& Caloi, V., 2008, accepted for
  publication in MNRAS (arXiv:0807.4233)
\bibitem[()]{Glatt08a} Glatt, K., et al. 2008a, AJ, 136, 1703
\bibitem[()]{Glatt08b} Glatt, K., et al. 2008b, AJ, 135, 1106
\bibitem[()]{Gratton01} Gratton, R. \ G., et al. 2001, A\&A, 369, 87
\bibitem[()]{Gratton04} Gratton, R. \ G., Sneden, C., \& Carretta,
  E. 2004, ARA\&A, 42, 385
\bibitem[()]{Mackey07} Mackey, A. \ D., \& Broby Nielsen, P.\ 2007,
  MNRAS, 369, 921 (M07)
\bibitem[()]{Mackey08} Mackey, A. \ D., Broby Nielsen, P., Ferguson,
  M. \ N. \& Richardson, J. \ C. 2008, \apj, 681, 17L (M08)
\bibitem[()]{Marino08} Marino, A. \ F., Villanova, S., Piotto, G.,
  Milone, A. \ P., Momany, Y., Bedin, L. \ R., \& Medling,
  A. \ M., 2008, accepted for pubblication in A\&A (arXiv:0808.1414)
\bibitem[()]{Milone08a} Milone, A. \ P.\ et al. 2008, \apj, 673, 241
\bibitem[()]{Milone08b} Milone, A. \ P., Piotto, G., Bedin, L. \ R. \&
  Sarajedini, A. 2008, in XXI Century Challenges for
  Stellar Evolution, Memorie della Societa Astronomica Italiana,
  vol. 79/2, eds: S. Cassisi, M. Salaris (arXiv:0801.3177)
\bibitem[()]{Mucciarelli07} Mucciarelli, A. Origlia, L., \& Ferraro,  F. \ R. 2007, AJ, 134, 1813
\bibitem[]{Pietrinferni04} Pietrinferni, A., Cassisi, S., Salaris, M., Castelli, F. \ 2004, \apj, 612, 168
\bibitem[]{Pietrinferni06} Pietrinferni, A., Cassisi, S., Salaris, M., Castelli, F. \ 2006, \apj, 642, 797
\bibitem[()]{Piotto05} Piotto, G.\ et al. 2005, \apj, 621, 777
\bibitem[()]{Piotto07} Piotto, G.\ et al. 2007, \apj, 661, L53
\bibitem[()]{Piotto08a} Piotto, G.\ 2008, in XXI Century Challenges for
  Stellar Evolution, Memorie della Societa Astronomica Italiana,
  vol. 79/2, eds: S. Cassisi, M. Salaris (arXiv:0801.3175)
\bibitem[()]{Piotto08b} Piotto, G.\ et al. in preparation
\bibitem[()]{Salpeter55} Salpeter, E. 1955, \apj, 121, 161
\bibitem[()]{Sarajedini08} Sarajedini, A.\ et al. 2007, AJ, 133, 290
\bibitem[()]{Sollima07} Sollima, A., Ferraro, F. \ R., Bellazzini, M.,
  Origlia, L., Straniero, O., \& Pancino, E. \ 2007, \apj, 654, 915
\bibitem[()]{Sirianni05} Sirianni, M.\ et al. 2005, PASP, 117, 1049
\bibitem[()]{Ventura01} Ventura, P., D'Antona, F., Mazzitelli, I., \&
  Gratton, R., 2001, \apj, 550, 65
\bibitem[()]{Villanova07} Villanova, S.\ et al. 2007, \apj, 663, 296
\bibitem[()]{Yong08} Yong, D., Grundahal, F., Johnson, J. \ A., \&
  Asplund, M. \ 2008, (arXiv:0806.0187v1)
\end{thebibliography}

%__________________________________________________________________
%%%%
\end{document}